\documentclass[a4paper,11pt]{article}
\pdfoutput=1 
\usepackage{jcappub} 
\usepackage{afterpage,booktabs,multirow} 
\usepackage{graphicx,subfig} 
\usepackage{amssymb,amsmath,bm} 
\usepackage[usenames,dvipsnames]{xcolor}
\usepackage[capitalise]{cleveref}
\usepackage{expl3,xstring,xparse} 
\usepackage{dcolumn}
\usepackage{bm}
\usepackage{relsize}
\usepackage{color}

\DeclareMathOperator{\BetaFac}{\kappa}
\DeclareMathOperator{\period}{.}
\DeclareMathOperator{\comma}{,}

\title{\boldmath COLA with massive neutrinos}
\author{Bill S. Wright,}
\author{Hans A. Winther,}
\author{and Kazuya Koyama}
\affiliation{Institute of Cosmology \& Gravitation, University of Portsmouth, Portsmouth, Hampshire, PO1 3FX, UK}
\emailAdd{bill.wright@port.ac.uk}

\date{\today}

\abstract{The effect of massive neutrinos on the growth of cold dark matter perturbations acts as a scale-dependent Newton's constant and leads to scale-dependent growth factors just as we often find in models of gravity beyond General Relativity. We show how to compute growth factors for $\Lambda$CDM and general modified gravity cosmologies combined with massive neutrinos in Lagrangian perturbation theory for use in COLA and extensions thereof. We implement this together with the grid-based massive neutrino method of Brandbyge and Hannestad in \texttt{MG-PICOLA} and compare COLA simulations to full {\it N}-body simulations of $\Lambda$CDM and $f(R)$ gravity with massive neutrinos. Our implementation is computationally cheap if the underlying cosmology already has scale-dependent growth factors and it is shown to be able to produce results that match {\it N}-body to percent level accuracy for both the total and CDM matter power-spectra up to $k\lesssim 1 h/$Mpc.}

\begin{document}
\maketitle

\section{Introduction}

Observations of neutrino flavour oscillations \cite{MassiveNeutrinos1, MassiveNeutrinos2} demand that at least two of the neutrino states are massive \cite{Pontecorvo1957}. While the absolute mass of each of the three mass eigenstates is unknown, there are strong constraints on the difference in mass between the states which implies that at least one of them has a mass greater than $\mathtt{\sim}$0.06 eV.

Massive neutrinos are known to affect the formation of structure in the Universe by suppressing structure formation at small physical scales \cite{Bondetal1980,2016PhRvD..94h3522G,2016PDU....13...77C,2015JCAP...02..045P,2014MNRAS.444.3501B,2013MNRAS.436.2038Z}. This is in contrast to the effect of modified gravity models, which tend to enhance structure formation. With many current observations providing strong constraints on deviations from the $\Lambda$CDM ($\Lambda$-Cold-Dark-Matter) cosmology \cite{Planck2015MG, Heavens2017}, the opposing effects of massive neutrinos and modified gravity offer a potential alternative to $\Lambda$CDM when combined \cite{2013PhRvL.110l1302M,2013PhRvD..88j3523H,Baldi2014,2015PhRvD..91f3524H,2010PThPh.124..541M,Bellomo2017,2017PhRvD..95f3502A}.

Multiple methods of including the effects of massive neutrinos in {\it N}-body simulations have been presented and used to make predictions for cosmological observables \citep{2009JCAP...05..002B,2010JCAP...09..014B,2012MNRAS.420.2551B,2013MNRAS.428.3375A,2014JCAP...03..011V,2014JCAP...02..049C,2013JCAP...12..012C,2012JCAP...02..045H,2016arXiv160503829H,2015JCAP...07..043C,2016JCAP...11..015B}. High-resolution {\it N}-body simulations are still the only game in town if one wants to produce simulations that are accurate over a wide range of scales; however, they have the downside of being very computationally expensive to perform. For the next generation of galaxy surveys such as Euclid \cite{Euclid}, eBOSS \cite{2016AJ....151...44D}, DESI \cite{DESI}, WFIRST \cite{WFIRST}, LSST \cite{LSST}, and SKA \cite{SKA}, there is a need to develop large ensembles of galaxy mocks that include the effects of massive neutrinos in order to model the observables and their covariances. For such purposes faster methods are needed (see e.g. \citep{1996ApJS..103....1B,2002MNRAS.329..629S,2013MNRAS.428.1036M,2015MNRAS.447..437M,2014MNRAS.437.2594W,2014MNRAS.439L..21K,2015MNRAS.450.1856A,OriginalCOLA,2016arXiv161206469V,2015arXiv150207751T,2016MNRAS.459.2118K,2015JCAP...06..015L,2016MNRAS.459.2327I,PINOCCHIO1,PINOCCHIO2} and \cite{2017arXiv170400920M} for a recent comparison of some of these methods).

In previous work \cite{Winther2017}, we presented a code {\tt{MG-PICOLA}}\footnote{The code can be found at https://github.com/HAWinther/MG-PICOLA-PUBLIC while the original {\tt{L-PICOLA}} code on which it is based can be found at https://github.com/CullanHowlett/l-picola} to perform fast, approximate numerical simulations of structure formation in models that have scale-dependent growth using the COmoving Lagrangian Acceleration (COLA) approach \cite{OriginalCOLA} (see also \cite{2016arXiv161206469V} for a similar approach). The COLA approach differs from traditional {\it N}-body simulations in that we solve for the perturbations about the particle trajectories predicted by Lagrangian perturbation theory. This allows us to take large {\it N}-body time-steps, thus saving a lot of computations (COLA is generally 2-3 orders of magnitude faster than {\it N}-body), while at the same time keeping accuracy on the largest scales, making the COLA approach well suited for producing large ensembles of galaxy mocks. In this paper we demonstrate how to extend this method to incorporate both modified gravity and massive neutrinos simultaneously in a quasi-nonlinear manner. The method is quasi-nonlinear because we only consider a linear neutrino density perturbation, such that the total density matter perturbation is defined $\delta_{\rm m} = ( \overline{\rho}_{\rm cb}\delta_{\rm cb} + \overline{\rho}_{\nu}\delta^{(1)}_{\nu} )/\overline{\rho}_{\rm m}$ where we track non-linearities only in $\delta_{\rm cb}$. Here the subscripts `m', `cb', and `$\nu$' represent the total matter, cold dark matter + baryon, and massive neutrino components respectively. $f_{\rm cb}$ and $f_{\nu}$ are the fractional contributions of those components to the total matter density parameter $\Omega_{\rm m}$. The superscript numbers refer to the order in the perturbative expansion.  We discuss the validity of only considering the linear neutrino density perturbation in Appendix \ref{sect:compnu}. We note that other fast, approximate methods can also handle massive neutrinos, namely PINOCCHIO (PINpointing Orbit-Crossing Collapsed HIerarchical Objects) \cite{PINOCCHIO1,PINOCCHIO2}.

The paper is organised as follows. In Section~\ref{sec:1storder} we recap how the first order growth factor is calculated in $\Lambda$CDM without massive neutrinos (Section~\ref{ssec:1storderLCDM}), modified gravity without massive neutrinos (Section~\ref{ssec:1storderMG}), and $\Lambda$CDM with massive neutrinos (Section~\ref{ssec:1storderLCDMmnu}), before explaining our extension for cosmologies with both modified gravity and massive gravity in Section~\ref{ssec:1storderMGmnu} and presenting the results. In Section~\ref{sec:2ndorder} we repeat this for the second order growth factors. In Section~\ref{sec:COLA} we describe how these growth factors are implemented into the COLA approach. We present and discuss the outputs of our extended code in Section~\ref{sec:Results} and then conclude in Section~\ref{sec:Conclusion}.

\section{First order growth factors}\label{sec:1storder}

\subsection{$\Lambda$CDM without massive neutrinos} \label{ssec:1storderLCDM}

In Lagrangian perturbation theory (LPT; see \cite{1996dmu..conf..565B} for a review) the position of a particle of species 's' ${x_{\mathrm{s}}}_{i}$ is written in terms of its initial position ${q_{\mathrm{s}}}_{i}$ and a displacement field ${\Psi_{\mathrm{s}}}_{i}$ as ${x_{\mathrm{s}}}_{i} = {q_{\mathrm{s}}}_{i} + {\Psi_{\mathrm{s}}}_{i}(\vec{q},\tau)$. Here $\tau$ is defined such that $d\tau = dt/a^2$. We expand the displacement field in a perturbation series
\begin{align}
{\Psi_{\mathrm{s}}}_{i} = \epsilon{\Psi_{\mathrm{s}}}_{i}^{(1)} + \epsilon^2{\Psi_{\mathrm{s}}}_{i}^{(2)} + \ldots\,\comma
\end{align} 
and by assuming ${\Psi_{\mathrm{s}}}_{i}$ is curl-free we can write ${\Psi_{\mathrm{s}}}_{i}^{(n)} = {\nabla_q}_i\phi^{(n)}_{\rm s}$ where $\phi^{(n)}_{\rm s}$ is a scalar field. We expand the cold dark matter (CDM) + baryon density contrast in a perturbation series, and use the Jacobian of the transformation between ${x_{\mathrm{cb}}}_{i}$ and ${q_{\mathrm{cb}}}_{i}$ to write
\begin{align}
\delta_{\rm{cb}} = \left|\frac{\partial (x,y,z)}{\partial (q_x,q_y,q_z)}\right|^{-1} - 1 = \epsilon\delta^{(1)}_{\rm{cb}} + \epsilon^2\delta^{(2)}_{\rm{cb}} + \ldots\, \period
\end{align}
Thus the CDM+baryon density contrast can be written in terms of the displacement-field order by order as
\begin{align}\label{eq:deltapsi1}
\delta^{(1)}_{\rm{cb}} &= - \Psi^{(1)}_{\mathrm{cb}\ i, i}\,\comma
\end{align}
\begin{align}\label{eq:deltapsi2}
\delta^{(2)}_{\rm{cb}} &= -\Psi^{(2)}_{\mathrm{cb}\ i, i} + \frac{1}{2}\left((\Psi^{(1)}_{\mathrm{cb}\ i, i})^2+(\Psi^{(1)}_{\mathrm{cb}\ i, j})^2\right)\comma
\end{align}
where $\Psi_{\mathrm{s}\ i, j} = \partial {\Psi_{\mathrm{s}}}_i / \partial {q}_j$. To first order, the Lagrangian equation of motion for the CDM+baryon component is
\begin{align}\label{eq:1storderEOM}
\frac{d^2}{d\tau^2}\Psi^{(1)}_{\mathrm{cb}\ i,i} = -{\nabla_x}^2\Phi_N\,\period
\end{align}
In $\Lambda$CDM the only matter in the late-time Universe sourcing the Newtonian potential $\Phi_N$ is CDM+baryons, such that we can write $\delta_{\rm{m}}=\delta_{\rm{cb}}$. To first order the Poisson equation is thus
\begin{align}\label{eq:Poisson1stLCDM}
{\nabla_x}^2\Phi_N = \BetaFac \delta_{\rm{cb}}^{(1)}\,\comma
\end{align}
where $\BetaFac = 4\pi G\overline{\rho}_{\rm{cb}} a^4 = \frac{3}{2}\Omega_{\rm{cb}} H_0^2 a$. The equations above combine to yield
\begin{align}
\left(\frac{d^2}{d\tau^2} - \BetaFac\right){\nabla_q}^2\phi^{(1)}_{\rm{cb}}(\vec{q}, \tau) = 0\comma
\end{align}
and ${\nabla_q}^2\phi^{(1)}_{\rm{cb}}(\vec{q},\tau_{\rm ini}) = -\delta^{(1)}_{\rm{cb}}(\vec{q},\tau_{\rm ini})$ follows from assessing Eq.~(\ref{eq:deltapsi1}) at $\tau_{\rm ini}$. We can separate out the time dependence with $\phi^{(1)}_{\rm{cb}}(\vec{q},\tau) = D_{1, \rm{cb}}(\tau)\phi^{(1)}_{\rm{cb}}(\vec{q},\tau_{\rm ini})$ where the growth factor $D_{1, \rm{cb}}$ only depends on time and satisfies the simple ODE
\begin{align}\label{eq:1LPTLCDM}
\frac{d^2D_{1, \rm{cb}}}{d\tau^2} - \BetaFac D_{1, \rm{cb}} = 0\period
\end{align}
The initial conditions are set such that $D_{1, \rm{cb}}(\tau_{\rm ini}) = 1$ and $\frac{dD_{1, \rm{cb}}(\tau_{\rm ini})}{d\tau} = \left.\left(\frac{1}{a}\frac{da}{d\tau}\right)\right|_{\tau=\tau_{\rm ini}}$ corresponding to the growing mode in a matter dominated universe (Einstein-de Sitter).

\subsection{Scale-dependent modified gravity without massive neutrinos} \label{ssec:1storderMG}

For modified gravity theories where the growth factor is scale-dependent the situation becomes a little more complicated than in $\Lambda$CDM. We will here consider a general, first order parametrisation of the gravitational potential in Fourier space
\begin{align}\label{eq:Poisson1stMG}
\mathcal{F}_q \left[ {\nabla_x}^2 \Phi_N(\vec{x},\tau) \right] = \BetaFac \mu_{\mathrm{MG}}(k,\tau)\mathcal{F}_q \left[ \delta_{\rm{cb}}^{(1)}(\vec{x},\tau) \right] \comma
\end{align}
where $\vec{k}$ is the Fourier coordinate with respect to coordinate $\vec{q}$, and $\mu_{\mathrm{MG}}$ is the effective Newton's constant for a modified gravity theory. The addition of the scale-dependent $\mu_{\mathrm{MG}}$ in the Poisson equation means that, unlike in $\Lambda$CDM, we can no longer separate time and space. However, we can instead separate time for each Fourier mode. Inserting Eq.~(\ref{eq:Poisson1stMG}) into the Fourier transform of Eq.~(\ref{eq:1storderEOM}) with respect to $q$ and using the Fourier transform of Eq.~(\ref{eq:deltapsi1}) with respect to $q$  yields
\begin{align}
\left(\frac{d^2}{d\tau^2} - \BetaFac\mu_{\mathrm{MG}}(k, \tau) \right)\phi^{(1)}_{\rm{cb}}(\vec{k},\tau) = 0\comma
\end{align}
which allows us to make the split $\phi^{(1)}_{\rm{cb}}(\vec{k},\tau) = D_{1, \rm{cb}}(k,\tau)\phi^{(1)}_{\rm{cb}}(\vec{k},\tau_{\rm ini})$ where the growth factor satisfies
\begin{align}\label{eq:1LPTMG}
\frac{d^2D_{1, \rm{cb}}(k, \tau)}{d\tau^2} - \BetaFac\mu_{\mathrm{MG}}(k,\tau) D_{1, \rm{cb}}(k, \tau) = 0\comma
\end{align}
with the initial field given by $k^2\phi^{(1)}_{\rm{cb}}(\vec{k},\tau_{\rm ini}) = \delta^{(1)}_{\rm{cb}}(\vec{k},\tau_{\rm ini})$. We assume that at early times the modified gravity effects are negligible such that the initial conditions are still those for an Einstein-de Sitter universe: $D_{1, \rm{cb}}(k, \tau_{\rm ini}) = 1$, $\frac{dD_{1, \rm{cb}}(k, \tau_{\rm ini})}{d\tau} = \left.\left(\frac{1}{a}\frac{da}{d\tau}\right)\right|_{\tau=\tau_{\rm ini}}$. This second order differential equation can be solved numerically at each set of $(k, \tau)$ values for a given model with specified $\mu_{\mathrm{MG}}(k, \tau)$. We list $\mu_{\mathrm{MG}}(k, \tau)$ formulae for different modified gravity models in Appendix \ref{sec:MGmodels}.

\subsection{$\Lambda$CDM with massive neutrinos} \label{ssec:1storderLCDMmnu}

In order to account for the additional effect of massive neutrinos on the first order growth factor $D_{1,  \rm{cb}}^{m_{\nu}=0}$ for matter perturbations in $\Lambda$CDM without massive neutrinos, we follow the practice of \cite{EisensteinHu1999} and use the following fitting formulae to calculate the first order growth factor of CDM+baryon perturbations $D_{1, \rm{cb}}$ and total matter perturbations $D_{1, \rm{cb}\nu}$ in cosmologies with massive neutrinos:
\begin{align}\label{eq:D1cb}
D_{1, \rm{cb}}(k, \tau) = \left[ 1 + {\left( \frac{D_{1,  \rm{cb}}^{m_{\nu}=0}(\tau)}{1+y_{fs}(\chi; f_{\nu{}})} \right)}^{0.7} \right]^{p_{\rm{cb}}/0.7} D_{1,  \rm{cb}}^{m_{\nu}=0}(\tau)^{1-p_{\rm{cb}}}\,\comma
\end{align}
\begin{align}\label{eq:D1cbnu}
D_{1, \rm{cb}\nu}(k, \tau) = \left[ f_{\rm{cb}}^{0.7/p_{\rm{cb}}} + {\left( \frac{D_{1,  \rm{cb}}^{m_{\nu}=0}(\tau)}{1+y_{fs}(\chi; f_{\nu{}})} \right)}^{0.7} \right]^{p_{\rm{cb}}/0.7} D_{1,  \rm{cb}}^{m_{\nu}=0}(\tau)^{1-p_{\rm{cb}}}\,\comma
\end{align}
where
\begin{align}
p_{\rm{cb}} \equiv \frac{1}{4} \left[ 5 - \sqrt[]{1+24f_{\rm{cb}}} \right] \geq 0\,\comma
\end{align}
\begin{align}
y_{fs}(\chi; f_{\nu}) = 17.2 f_{\nu} \left( 1 + 0.488f_{\nu}^{-7/6} \right) \left( N_{\nu} \chi/f_{\nu} \right)^2\,\comma
\end{align}
\begin{align}
\chi = \frac{k}{\mathrm{Mpc^{-1}}} \Theta_{2.7}^2(\Omega_0h^2)^{-1} = \frac{k}{19.0}(\Omega_0 H_0^2)^{-1/2}(1+z_{\rm eq})^{-1/2}\,\comma
\end{align}
and $f_i=\Omega_i/\Omega_{\mathrm{m}}$ is the density ratio for species $i$, $N_{\nu}$ is the number of massive neutrino species, and $\Theta_{2.7}$ is a measure of the CMB temperature at $z=0$ using $T_{\mathrm{CMB}}=2.7\Theta_{2.7}\mathrm{K}$.

This method allows us to insert $\mathrm{\Lambda}$CDM $D_{1,  \rm{cb}}^{m_{\nu}=0}$ values that we calculate by solving Eq.~(\ref{eq:1LPTLCDM}) into Eqs.~(\ref{eq:D1cb}, \ref{eq:D1cbnu}) to calculate $D_{1,  \rm{cb}}^{m_{\nu}}$ and $D_{1,  \rm{cb}\nu}^{m_{\nu}}$ which then include the effect of massive neutrinos at linear order. 

An alternative method for computing $D_{1,  \rm{cb}}$ values would be to use the output of a Boltzmann solver code, such as {\tt{CAMB}} \citep{CAMB}. In our implementation, we allow for both methods to be used as per the user's preference.

\begin{figure*}
\includegraphics[height=0.3515625\textheight]{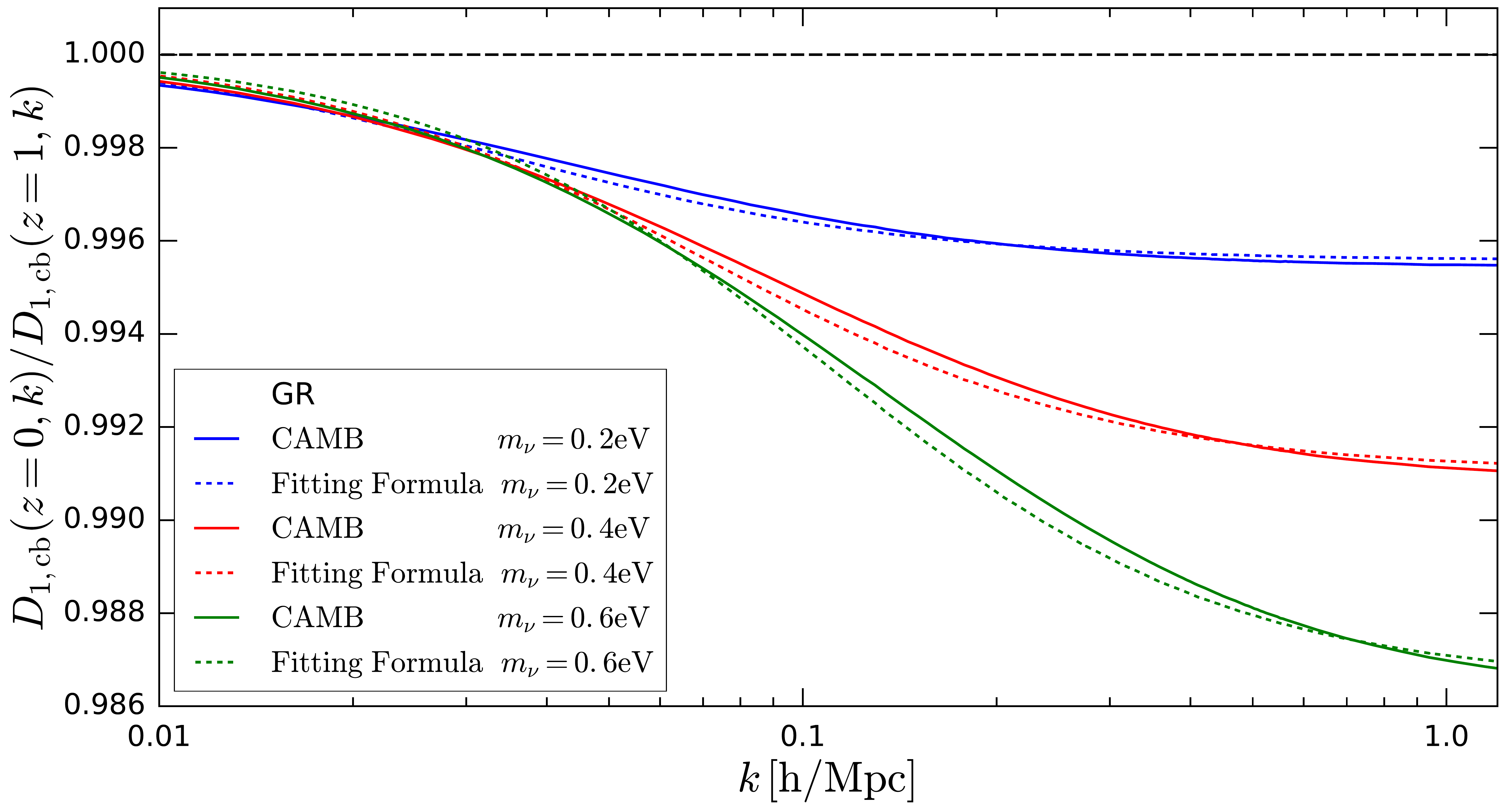}
\caption{Comparison between {\tt{CAMB}} and the fitting formula method Eq.~(\ref{eq:D1cb}) for the ratio of the first order CDM+baryon growth factor at $z=0$ to $z=1$ for a GR + massive neutrino cosmology with $m_{\nu}=[0.2, 0.4, 0.6]$ eV. The ratio has been normalised to the $\Lambda$CDM case without neutrinos, which is given by the horizontal dashed line.}
\label{fig:MGCAMBvFitvNum_grmnu}
\end{figure*}

Figure~\ref{fig:MGCAMBvFitvNum_grmnu} displays a comparison between the outputs of Eq.~(\ref{eq:D1cb}) and {\tt{CAMB}}. Specifically, the figure plots the ratio of the first order CDM+baryon growth factor at $z=0$ and $z=1$, $D_{1,  \rm{cb}}(k, z=0)/D_{1,  \rm{cb}}(k, z=1)$. The comparison is done for three GR+massive neutrino cosmologies with $m_{\nu}=[0.2, 0.4, 0.6]$ eV. The ratio has also been normalised to the $\Lambda$CDM case without massive neutrinos. Figure~\ref{fig:MGCAMBvFitvNum_grmnu} shows that the output of Eq. (\ref{eq:D1cb}) matches ${\tt{CAMB}}$ to an accuracy of $< 1\%$ for neutrino masses $m_{\nu} \lesssim 0.6$ eV up to $k=1.0\ h/\mathrm{Mpc}$.

\subsection{Scale-dependent modified gravity with massive neutrinos} \label{ssec:1storderMGmnu}

Eisenstein and Hu originally only considered $\mathrm{\Lambda}$CDM $D_{1,  \rm{cb}}^{m_{\nu}=0}$ values as input to Eqs. (\ref{eq:D1cb}, \ref{eq:D1cbnu}). In this work we extend their idea by using $D_{1,  \rm{cb}}^{m_{\nu}=0}$ values for modified gravity cosmologies without massive neutrinos as input instead, these having been calculated numerically by solving Eq. (\ref{eq:1LPTMG}). This adds the effect of massive neutrinos to the modified gravity models, enabling us to compute growth factor values for modified gravity + massive neutrino (MG+$m_{\nu}$) cosmologies at linear order.

\begin{figure*}
\includegraphics[height=0.3515625\textheight]{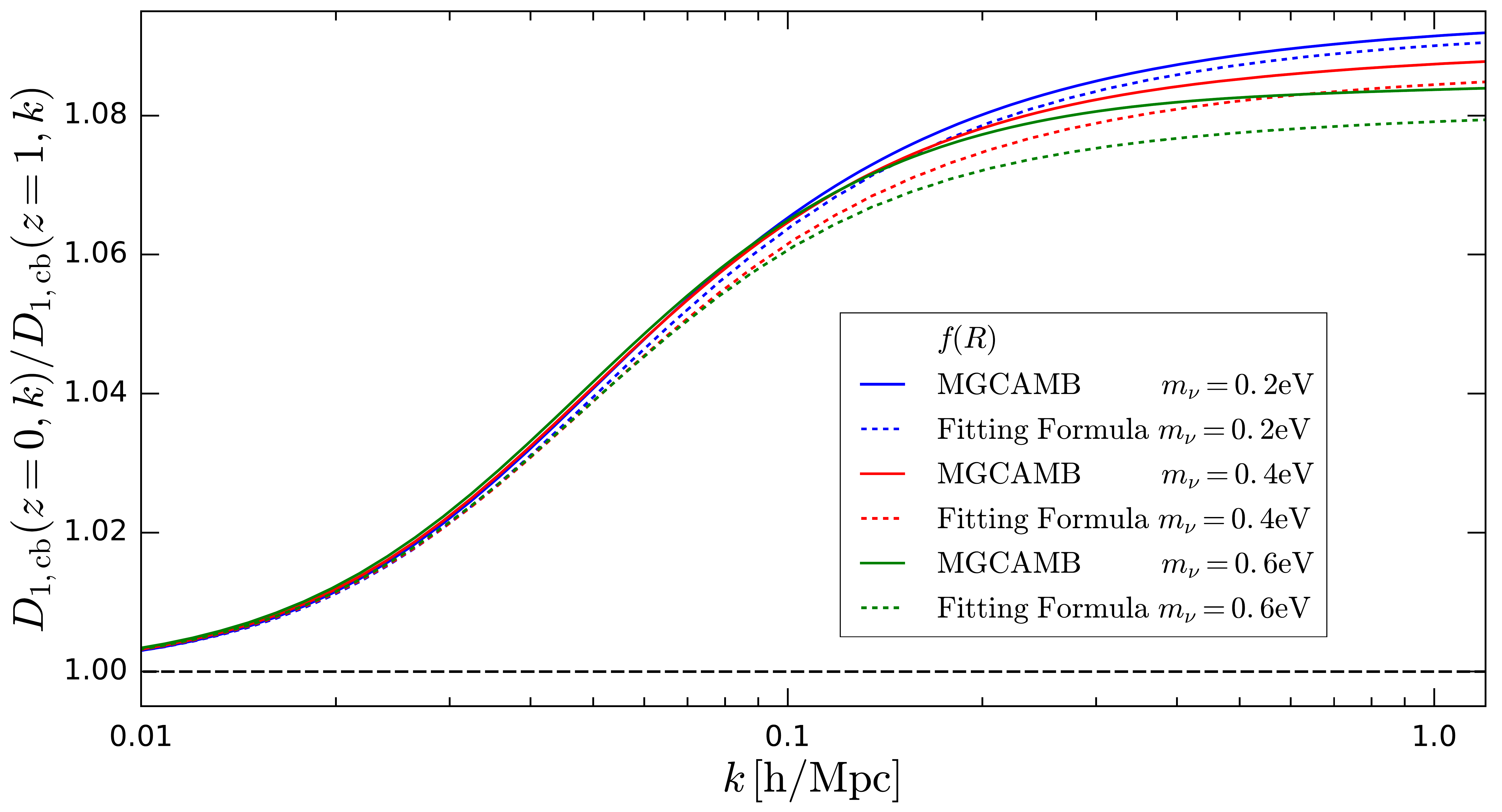}
\caption{Comparison between {\tt{MGCAMB}} and the fitting formula method Eq.~(\ref{eq:D1cb}) for the ratio of the first order CDM+baryon growth factor at $z=0$ to $z=1$ for a $f(R)$ + massive neutrino cosmology with $|f_{R0}| = 10^{-4}$ and $m_{\nu}=[0.2, 0.4, 0.6]$ eV. The ratio has been normalised to the $\Lambda$CDM case without massive neutrinos, which is given by the horizontal dashed line.}
\label{fig:MGCAMBvFitvNum_fofrmnu}
\end{figure*}

\begin{figure*}
\includegraphics[height=0.3515625\textheight]{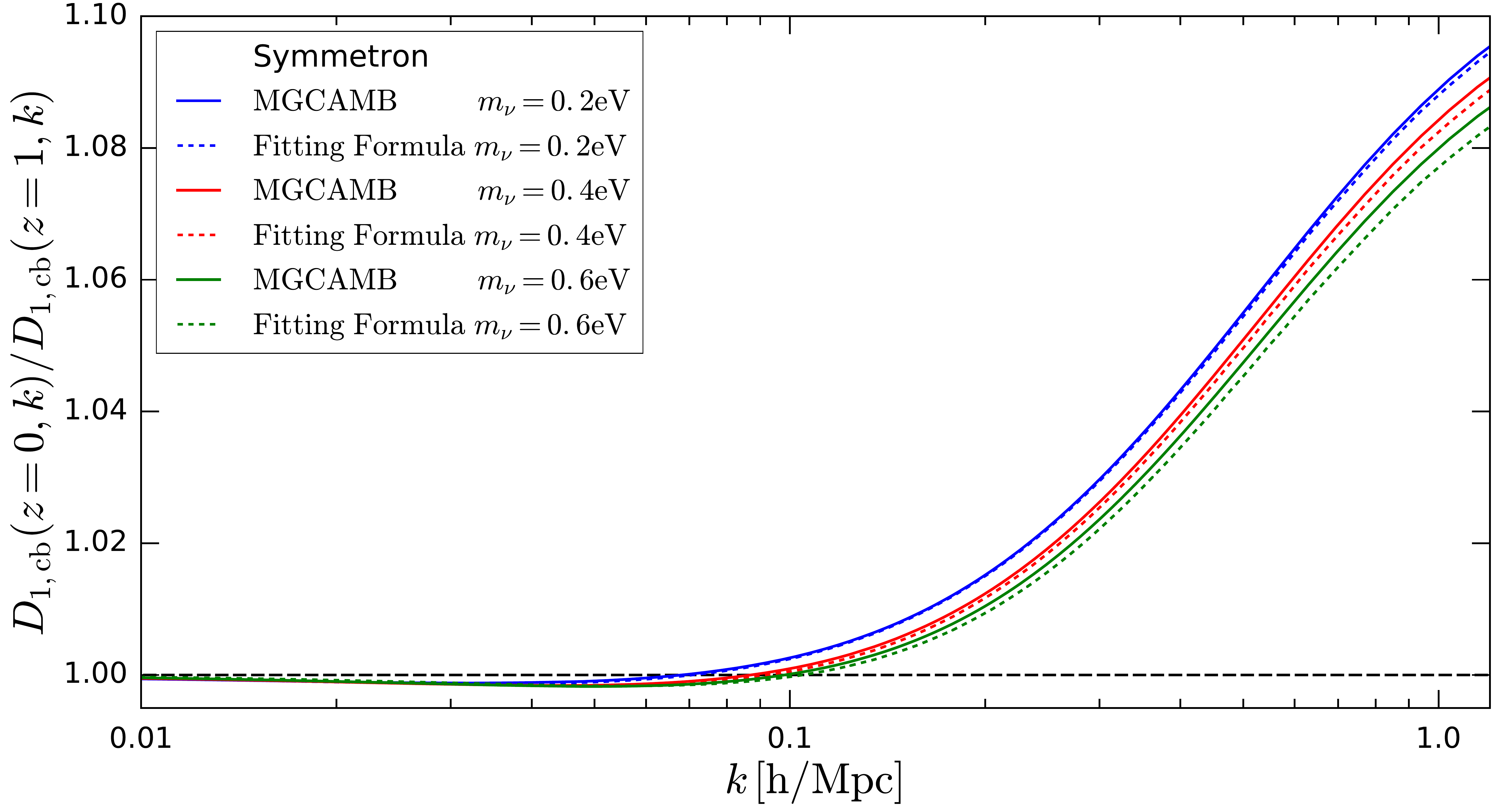}
\caption{Comparison between {\tt{MGCAMB}} and the fitting formula method Eq.~(\ref{eq:D1cb}) for the ratio of the first order CDM+baryon growth factor at $z=0$ to $z=1$ for a symmetron + massive neutrino cosmology with $\beta_{\star}=1$, $a_{\star}=0.5$, $\xi_{\star}=1/2998$, and $m_{\nu}=[0.2, 0.4, 0.6]$ eV. The ratio has been normalised to the $\Lambda$CDM case without massive neutrinos, which is given by the horizontal dashed line.}
\label{fig:MGCAMBvFitvNum_symmnu}
\end{figure*}

\begin{figure*}
\includegraphics[height=0.3515625\textheight]{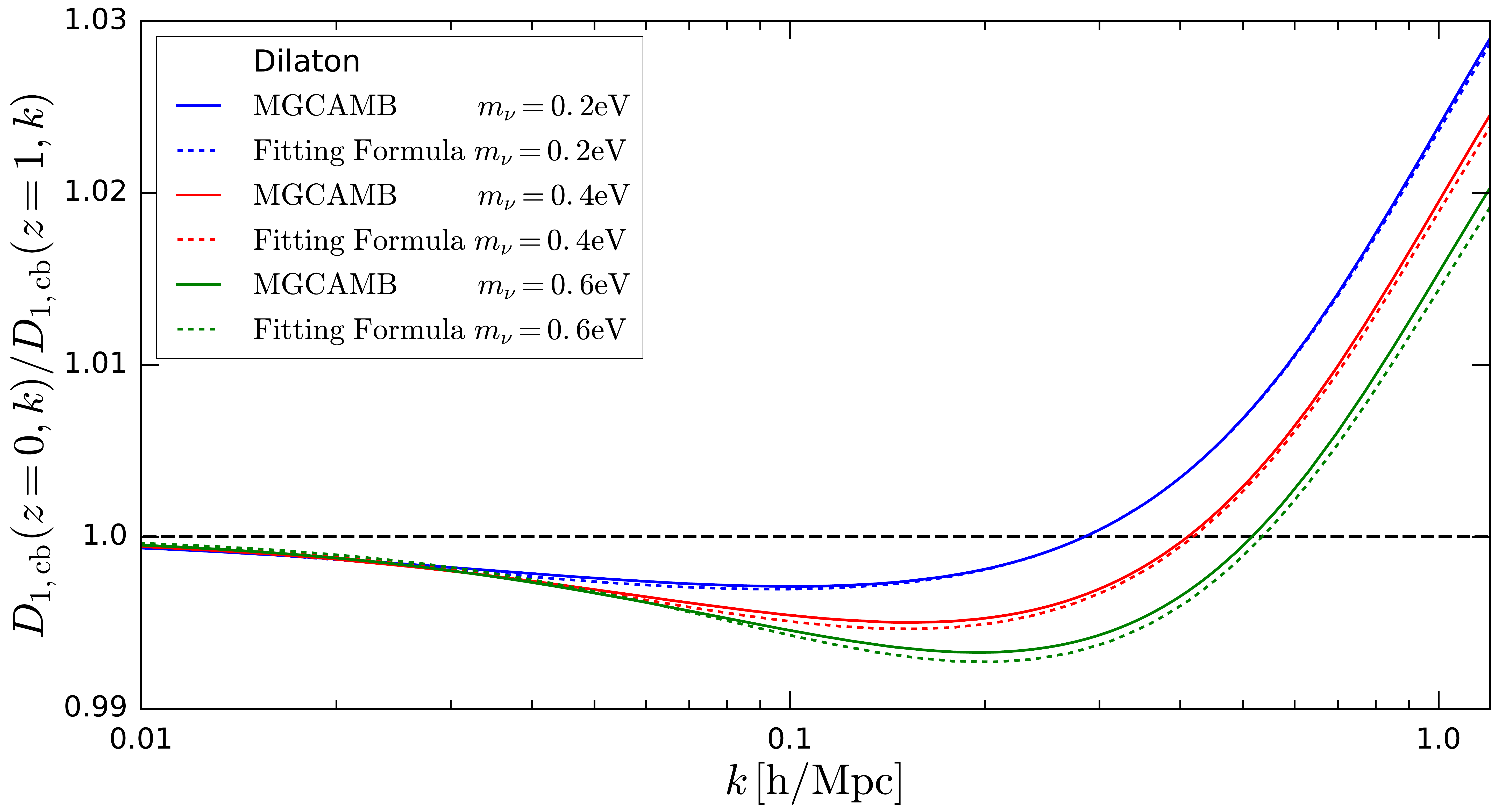}
\caption{Comparison between {\tt{MGCAMB}} and the fitting formula method Eq.~(\ref{eq:D1cb}) for the ratio of the first order CDM+baryon growth factor at $z=0$ to $z=1$ for a dilaton + massive neutrino cosmology with $\beta_{0}=0.41$, $\xi_{0}=1/2998$, $S=0.24$, $R=1$, and $m_{\nu}=[0.2, 0.4, 0.6]$ eV. The ratio has been normalised to the $\Lambda$CDM case without massive neutrinos, which is given by the horizontal dashed line.}
\label{fig:MGCAMBvFitvNum_dilmnu}
\end{figure*}

To test the fitting formula method for growth factors in MG+$m_{\nu}$ cosmologies, we first verified that our extension to the fitting formula method gives accurate values of $D_{1, \rm{cb}}$ and $D_{1, \rm{cb}\nu}$ by comparison with the output from {\tt{MGCAMB}}, which is an extension of {\tt{CAMB}} for modified gravity models \citep{MGCAMB1,MGCAMB2}. These comparisons can be seen in Figures~\ref{fig:MGCAMBvFitvNum_fofrmnu}-\ref{fig:MGCAMBvFitvNum_dilmnu} for $D_{1, \rm{cb}}$ in the Hu-Sawicki $f(R)$, symmetron, and dilaton modified gravity models. As in Figure~\ref{fig:MGCAMBvFitvNum_grmnu}, the plots show the ratio of the first order CDM+baryon growth factor at $z=0$ and $z=1$, $D_{1,  \rm{cb}}(k, z=0)/D_{1,  \rm{cb}}(k, z=1)$, the ratios have been normalised to the $\Lambda$CDM case without massive neutrinos, and the comparison is made for three different neutrino masses $m_{\nu}=[0.2, 0.4, 0.6]$ eV. Figure~\ref{fig:MGCAMBvFitvNum_fofrmnu} shows that the ability of the fitting formula to recover {\tt{MGCAMB}} first order growth values for $f(R)+m_{\nu}$ cosmologies decreases as $m_{\nu}$ increases. However, for the F4 model of $f(R)$ gravity (where $|f_{R0}| = 10^{-4}$) that we consider in Figure~\ref{fig:MGCAMBvFitvNum_fofrmnu} the output of Eq.~(\ref{eq:D1cb}) matches that of {\tt{MGCAMB}} to an accuracy of $< 1\%$ up to $k=1.0\ h/\mathrm{Mpc}$ even for $m_{\nu}=0.6$ eV. Similarly, for the values of parameters we have considered here, Eq.~(\ref{eq:D1cb}) matches {\tt{MGCAMB}} to an accuracy of $< 1\%$ up to $k=1.0\ h/\mathrm{Mpc}$ for both the symmetron and dilaton cosmologies with neutrino masses $m_{\nu} \lesssim 0.6$ eV.

\begin{figure*}
\includegraphics[height=0.36\textheight]{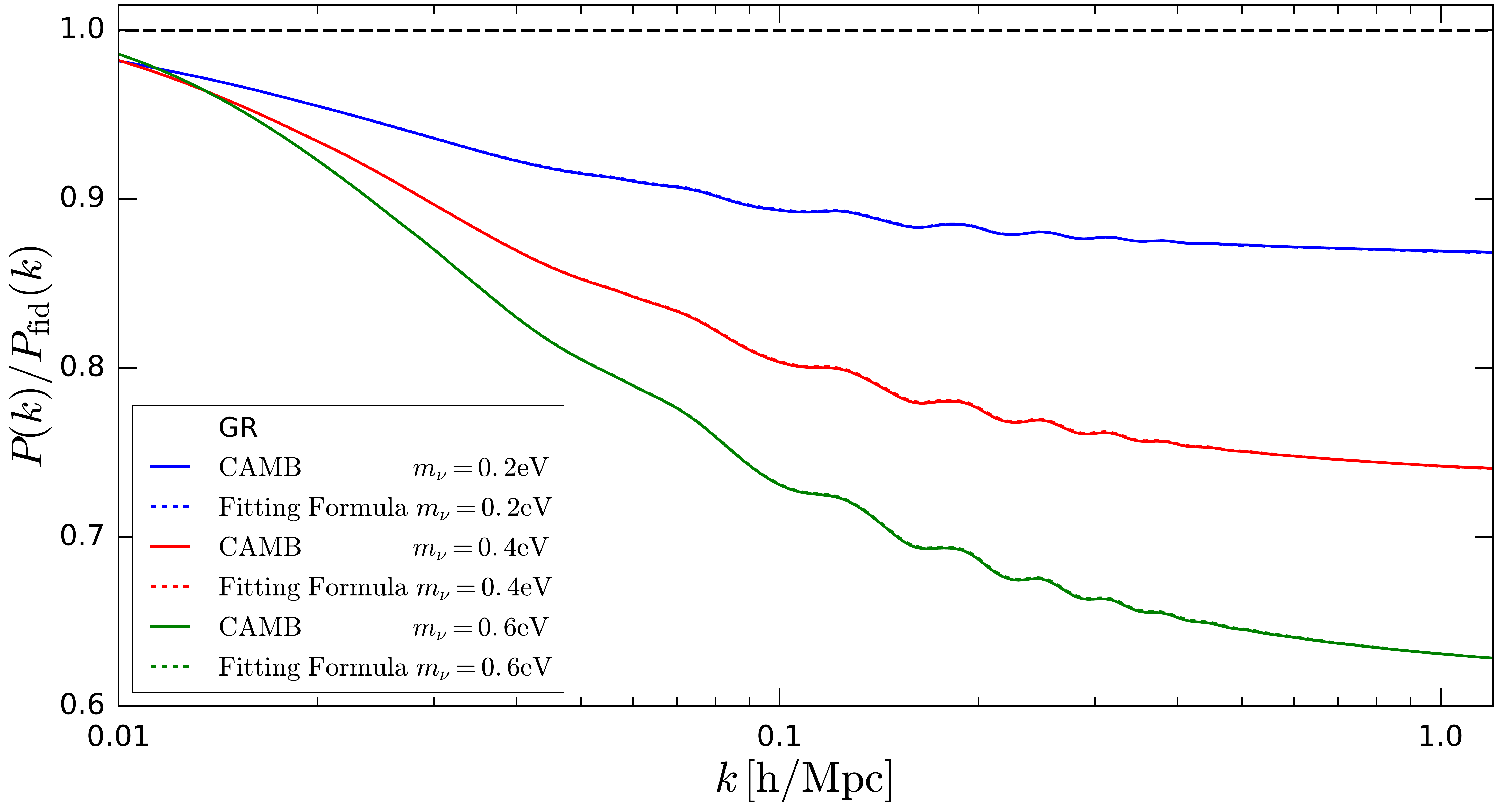}
\caption{The linear total matter power-spectrum at $z=1.3$ for a GR + massive neutrino cosmology with $m_{\nu}=[0.2, 0.4, 0.6]$ eV, calculated using both {\tt{CAMB}} and the fitting formula method Eq.~(\ref{eq:D1cb}). This power-spectrum is normalised to the fiducial $\Lambda$CDM case without massive neutrinos, which is shown by the horizontal dashed line.}
\label{fig:Pk_GR_z13}
\end{figure*}

\begin{figure*}
\includegraphics[height=0.36\textheight]{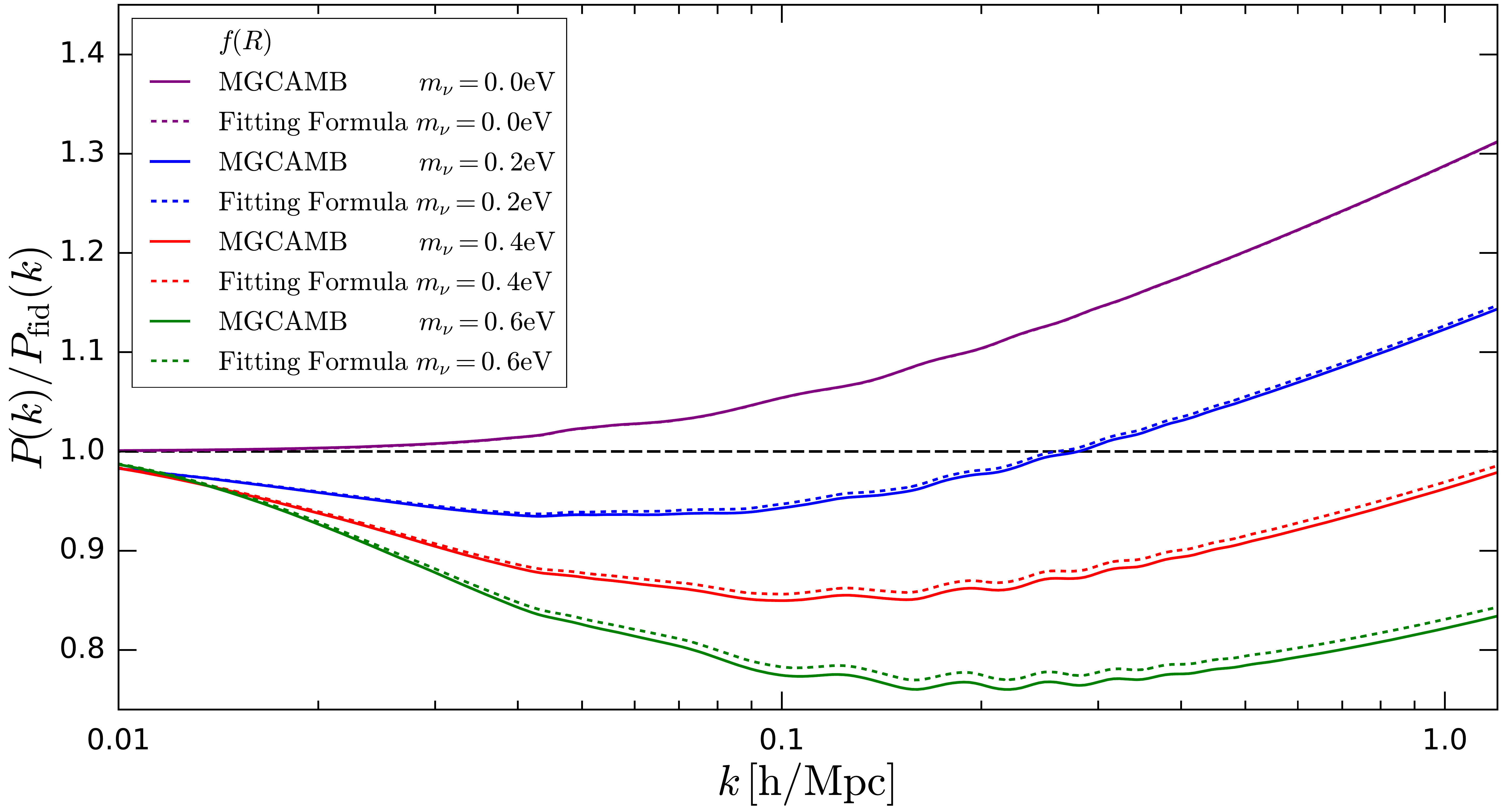}
\caption{The linear total matter power-spectrum at $z=1.3$ for a $f(R)$ + massive neutrino cosmology with $|f_{R0}| = 10^{-4}$ and $m_{\nu}=[0.0, 0.2, 0.4, 0.6]$ eV, calculated using both {\tt{MGCAMB}} and the fitting formula method Eq.~(\ref{eq:D1cb}). This power-spectrum is normalised to the fiducial $\Lambda$CDM case without massive neutrinos, which is shown by the horizontal dashed line.}
\label{fig:Pk_fofr_z13}
\end{figure*}

\begin{figure*}
\includegraphics[height=0.3515625\textheight]{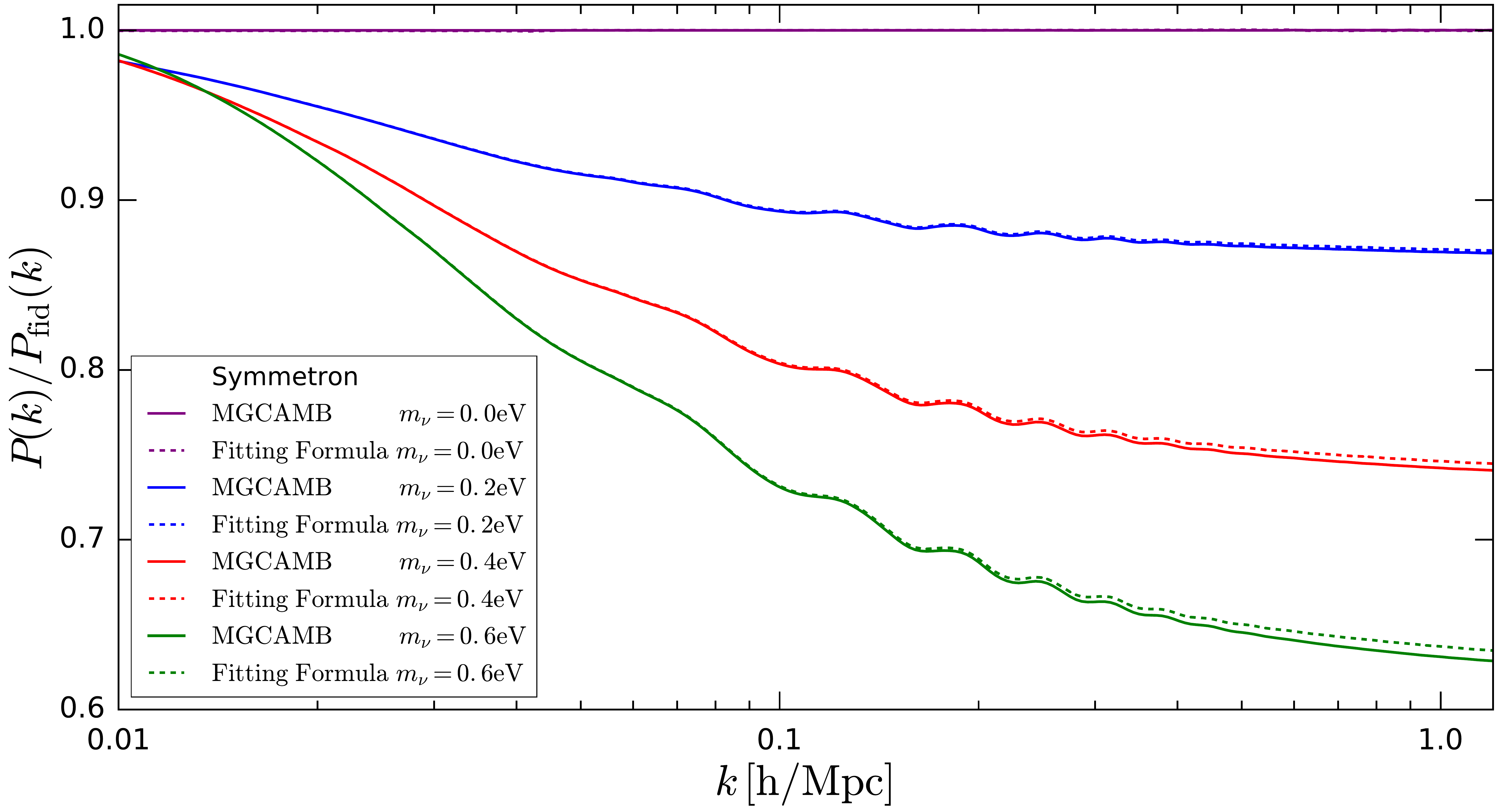}
\caption{The linear total matter power-spectrum at $z=1.3$ for a symmetron + massive neutrino cosmology with $\beta_{\star}=1$, $a_{\star}=0.5$, $\xi_{\star}=1/2998$, and $m_{\nu}=[0.0, 0.2, 0.4, 0.6]$ eV, calculated using both {\tt{MGCAMB}} and the fitting formula method Eq.~(\ref{eq:D1cb}). This power-spectrum is normalised to the fiducial $\Lambda$CDM case without massive neutrinos, which is shown by the horizontal dashed line.}
\label{fig:Pk_sym_z13}
\end{figure*}

\begin{figure*}
\includegraphics[height=0.3515625\textheight]{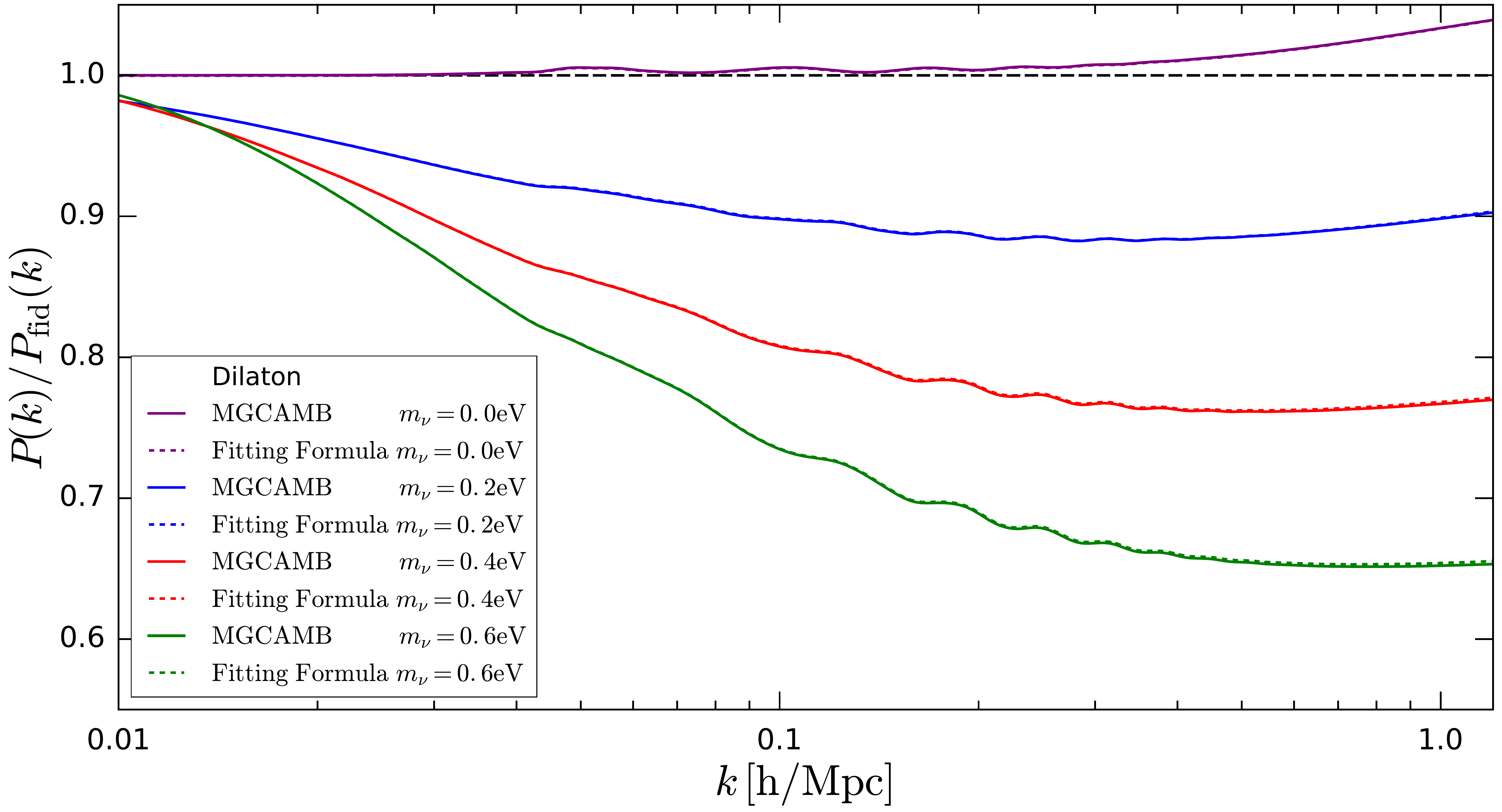}
\caption{The linear total matter power-spectrum at $z=1.3$ for a dilaton + massive neutrino cosmology with $\beta_{0}=0.41$, $\xi_{0}=1/2998$, $S=0.24$, $R=1$, and $m_{\nu}=[0.0, 0.2, 0.4, 0.6]$ eV, calculated using both {\tt{MGCAMB}} and the fitting formula method Eq.~(\ref{eq:D1cb}). This power-spectrum is normalised to the fiducial $\Lambda$CDM case without massive neutrinos, which is shown by the horizontal dashed line.}
\label{fig:Pk_dil_z13}
\end{figure*}

We also wanted to test whether the growth factors calculated using this method could be used to accurately `backscale' the linear total matter (CDM+baryon+massive neutrino) power-spectra at $z=0$ so that they closely matched the linear total matter power-spectra output at earlier $z$ by {\tt{MGCAMB}} directly. This was done using the relationship
\begin{align}
P_{\rm{cb}\nu}(k, z) = \left[\frac{D_{1, {\rm{cb}\nu}}(k, z)}{D_{1, {\rm{cb}\nu}}(k, z=0)}\right]^2 P_{\rm{cb}\nu}(k, z=0)\,\period
\end{align}
In Figures \ref{fig:Pk_GR_z13}-\ref{fig:Pk_dil_z13}, we display the resulting backscaled total matter linear power-spectra for the GR, $f(R)$, symmetron, and dilaton gravity models with $m_{\nu}$=[0, 0.2, 0.4, 0.6]eV at $z=1.3$. As for the first order growth factors, we find that, for the values of the model parameters considered, using the fitting formula method to backscale the $z=0$ linear total matter power-spectrum to $z=1.3$ recovers the same result as is output by {\tt{MGCAMB}} directly at $z=1.3$ to an accuracy of $< 1\%$ up to $k=1.0\ h/\mathrm{Mpc}$ for each of the cosmologies with neutrino masses $m_{\nu} \lesssim 0.6$ eV.

\section{Second order growth factors} \label{sec:2ndorder}

\subsection{$\Lambda$CDM without massive neutrinos} \label{ssec:2ndorderLCDM}

To second order, the Lagrangian equation of motion for the CDM+baryon component is
\begin{align}\label{eq:2ndorderEOM}
\frac{d^2\Psi^{(2)}_{\mathrm{cb}\ i, i}}{d\tau^2} - \Psi^{(1)}_{\mathrm{cb}\ j, i}\frac{d^2\Psi^{(1)}_{\mathrm{cb}\ i, j}}{d\tau^2} = - {\nabla_x}^2\Phi_N\,\comma
\end{align}
and the second order Poisson equation is
\begin{align}\label{eq:Poisson2ndLCDM}
{\nabla_x}^2\Phi_N = \BetaFac \delta_{\rm{cb}}^{(2)}\,\period
\end{align}
Using $\Psi^{(n)}_{i}={\nabla_q}_i\phi^{(n)}$ and Eq.~(\ref{eq:deltapsi2}), the above two equations combine to yield
\begin{align}
\left(\frac{d^2}{d\tau^2} - \BetaFac\right){\nabla_q}^2\phi^{(2)}_{\rm{cb}} = - \frac{\BetaFac}{2}\left[ ({\nabla_q}^2\phi^{(1)}_{\rm{cb}})^2 - ({\nabla_q}_i{\nabla_q}_j\phi^{(1)}_{\rm{cb}})^2\right]\,\period
\end{align}
Again we can separate out the time dependency completely with $\phi^{(2)}_{\rm{cb}}(\vec{q},\tau) = D_{2, \rm{cb}}(\tau)\phi^{(2)}_{\rm{cb}}(\vec{q},\tau_{\rm ini})$ and write
\begin{align}\label{eq:2lptLCDM}
\frac{d^2D_{2, \rm{cb}}}{d\tau^2} - \BetaFac D_{2, \rm{cb}} = -\BetaFac D_{1, \rm{cb}}^2\,\period
\end{align}
For an Einstein-de Sitter Universe the physically relevant solution has $D_{2, \rm{cb}} = -\frac{3}{7}D_{1, \rm{cb}}^2$ so the initial conditions are taken to be $D_{2, \rm{cb}}(\tau_{\rm ini}) = -\frac{3}{7}$ and $\frac{dD_{2, \rm{cb}}(\tau_{\rm ini})}{d\tau} = -\frac{6}{7}\left.\left(\frac{1}{a}\frac{da}{d\tau}\right)\right|_{\tau=\tau_{\rm ini}}$.
The initial field $\phi^{(2)}_{\rm{cb}}(\vec{q},\tau_{\rm ini})$ satisfies
\begin{align}\label{eq:lcdm2lpt}
{\nabla_q}^2\phi^{(2)}_{\rm{cb}}(\vec{q},\tau_{\rm ini}) = \frac{1}{2}\left[\left({\nabla_q}^2\phi^{(1)}_{\rm{cb}}(\vec{q},\tau_{\rm ini})\right)^2 - \left({\nabla_q}_i{\nabla_q}_j\phi^{(1)}_{\rm{cb}}(\vec{q},\tau_{\rm ini})\right)^2\right]\,\period
\end{align}

\subsection{Scale-dependent modified gravity without massive neutrinos} \label{ssec:2ndorderMG}

In a theory where the growth factor is scale-dependent, there are several additional complications at second order. When taking the Fourier transform of Eq.~(\ref{eq:2ndorderEOM}) with respect to $q$ we will need an expression for $\mathcal{F}_q \left[ {\nabla_x}^2 {\Phi}_N(\vec{x}, \tau) \right]$ which will contain frame-lagging terms above linear order \cite{AvilesCota}. For the general case where the scale-dependence at n$^{\rm th}$ order is encapsulated in the effective Newton's constant $\mu^{(n)}(k, \tau)$ the Fourier transform with respect to $x$ up to second order is
\begin{multline}
\mathcal{F}_x \left[ {\nabla_x}^2 {\Phi}(\vec{x}, \tau) \right] = \BetaFac \mu^{(1)}(k_{\rm E}, \tau) \delta^{(1)}_{\rm{cb}}(\vec{k}_{\rm E}, \tau) + \BetaFac \mu^{(2)}(k_{\rm E}, \tau) \delta^{(2)}_{\rm{cb}}(\vec{k}_{\rm E}, \tau) 
\\ + a^4 H^2\int \frac{d^3\vec{k}_1d^3\vec{k}_2}{(2\pi)^3} \delta_D(\vec{k}-\vec{k_{12}})\gamma_2^{\rm E}(\vec{k}, \vec{k_1}, \vec{k_2}, \tau) \delta^{(1)}_{\rm{cb}}(\vec{k_1}, \tau) \delta^{(1)}_{\rm{cb}}(\vec{k_2}, \tau)\comma
\end{multline}
where $\vec{k}_{\rm E}$ is the Fourier coordinate with respect to $\vec{x}$, $\delta_{\rm{cb}}(\vec{k}_{\rm E}, \tau) = \mathcal{F}_x \left[ \delta_{\rm{cb}}(\vec{x}, \tau) \right](\vec{k}_{\rm E})$ and $\vec{k}_{12} = \vec{k}_1 + \vec{k}_2$. We can show that the corresponding Fourier transform with respect to $q$ at second order for the specific case of scale-dependent modified gravity without massive neutrinos is \citep{BoseKoyama2016,AvilesCota,Winther2017}
\begin{multline}\label{eq:Poisson2ndMG}
\mathcal{F}_q \left[ {\nabla_x}^2 {\Phi}(\vec{x}, \tau) \right] = \BetaFac \mu_{\rm MG}(k, \tau) \tilde{\delta}^{(2)}_{\rm{cb}}(\vec{k}, \tau) \\ + a^4 H^2\int \frac{d^3\vec{k}_1d^3\vec{k}_2}{(2\pi)^3} \delta_D(\vec{k}-\vec{k_{12}})\gamma_2(\vec{k}, \vec{k_1}, \vec{k_2}, \tau) \delta^{(1)}_{\rm{cb}}(\vec{k_1}, \tau) \delta^{(1)}_{\rm{cb}}(\vec{k_2}, \tau) \,\comma
\end{multline}
where we have defined $\gamma_2 = \gamma_2^{\rm E} + \frac{3}{2}\Omega_{\rm m}(\tau)\left[ \mu_{\rm MG}(k, \tau) - \mu_{\rm MG}(k_1, \tau) \right] \frac{\vec{k}_1 \cdot \vec{k}_2}{k_2^2}$ and $\tilde{\delta}^{(2)}_{\rm{cb}}=\mathcal{F}_q \left[\delta^{(2)}_{\rm{cb}}(\vec{q}, \tau) \right](\vec{k}) + \mathcal{O}(\epsilon^3)$. There is also a frame-lagging effect on $\gamma_2^{\rm E}$, but this would be a third order term so it is not included here.
\\
We write $\phi^{(2)}_{\rm{cb}}$ (in a way that will become clear later) as an expansion
\begin{align}\label{eq:psi2lptintegral}
\phi^{(2)}_{\rm{cb}}(\vec{k},\tau) = -\frac{1}{2k^2}\int\frac{{\rm d}^3\vec{k}_1{\rm d}^3\vec{k}_2}{(2\pi)^3}\delta_D(\vec{k}-\vec{k}_{12}) \delta^{(1)}_{\rm{cb}}(\vec{k_1},\tau_{\rm ini})\delta^{(1)}_{\rm{cb}}(\vec{k_2},\tau_{\rm ini})D_{2, \rm{cb}}(\vec{k},\vec{k_1},\vec{k_2},\tau)\,\comma
\end{align}
where $\delta^{(1)}_{\rm{cb}}(\vec{k}, \tau_{\rm ini})$ corresponds to the initial density field. With $\phi_{\rm cb}^{(2)}$ written in this form, substituting Eq.~(\ref{eq:Poisson2ndMG}) into the Fourier space version of Eq.~(\ref{eq:2ndorderEOM}) leads to the second order differential equation for $D_{2, \mathrm{cb}}$
\begin{align} \label{eq:2LPTMGexact}
\left[ \frac{d^2}{d\tau^2} - \BetaFac \mu_{\rm MG}(k, \tau) \right] D_{2, \rm{cb}}(\vec{k},\vec{k_1},\vec{k_2},\tau) = - \BetaFac \mu_{\rm MG}(k, \tau) D_{1, \mathrm{cb}}(k_1, \tau) D_{1, \mathrm{cb}}(k_2, \tau) 
\nonumber \\ \times \left\{ 1 - \left[ \frac{2\mu_{\rm MG}(k_1, \tau) - \mu_{\rm MG}(k, \tau)}{\mu_{\rm MG}(k, \tau)} \right]\frac{ ( \vec{k}_1 \cdot \vec{k}_2 )^2}{k_1^2 k_2^2} \right.
\nonumber \\ \left.+ \frac{2a^4 H^2 \gamma_2(\vec{k}, \vec{k_1}, \vec{k_2}, \tau)}{\BetaFac \mu_{\rm MG}(k, \tau)} \right\} \comma
\end{align}
with initial conditions
\begin{align}
D_{2, \rm{cb}}(\vec{k}, \vec{k}_1, \vec{k}_2, \tau_{\rm ini}) &= -\frac{3}{7}\left(1 - \frac{(\vec{k_1}\cdot\vec{k_2})^2}{k_1^2k_2^2}\right)\,\comma\\
\frac{dD_{2, \rm{cb}}(\vec{k}, \vec{k}_1, \vec{k}_2, \tau_{\rm ini})}{d\tau} &=  -\frac{6}{7}\left(1 - \frac{(\vec{k_1}\cdot\vec{k_2})^2}{k_1^2k_2^2}\right)\left.\left(\frac{1}{a}\frac{da}{d\tau}\right)\right|_{\tau=\tau_{\rm ini}}\,\period
\end{align}
For most modified gravity models $\gamma_2$ is only a function of the wavenumber norms $k,k_1,k_2$ and the dot-product $\vec{k_1}\cdot \vec{k_2}$. The $\delta_D$ function in the integral for $D_2$ enforces $\vec{k_2} = \vec{k} - \vec{k_1}$, so solving Eq.~(\ref{eq:2LPTMGexact}) becomes a three-dimensional problem such that we only need to solve it for the combinations of $k$, $k_1$ and $\cos\theta \equiv \frac{\vec{k_1}\cdot \vec{k_2}}{k_1k_2}$ that correspond to a valid triangle in Fourier space.

Evaluating the integral in Eq.~(\ref{eq:psi2lptintegral}) at each time-step, without being able to rely on fast Fourier transforms, would ruin the speed of the COLA approach. As in \cite{Winther2017}, we therefore settle on an approximation for this term. We replace Eq.~(\ref{eq:psi2lptintegral}) with
\begin{align}\label{eq:phi2approx}
\phi^{(2)}_{\rm{cb}}(\vec{k},\tau) =& -\frac{\hat{D}_{2, \rm{cb}}(k,\tau)}{2k^2}&\nonumber \\ &\times \int\frac{{\rm d}^3\vec{k}_1{\rm d}^3\vec{k}_2}{(2\pi)^3}\delta_D(\vec{k}-\vec{k}_{12}) \delta^{(1)}_{\rm{cb}}(\vec{k_1},\tau_{\rm ini})\delta^{(1)}_{\rm{cb}}(\vec{k_2},\tau_{\rm ini})\left(1 - \frac{(\vec{k_1}\cdot\vec{k_2})^2}{k_1^2k_2^2}\right)\,\comma
\end{align}
and replace Eq.~(\ref{eq:2LPTMGexact}) with
\begin{align}\label{eq:2LPTMGapprox}
\frac{d^2\hat{D}_{2, \rm{cb}}}{d\tau^2} - \BetaFac \mu_{\mathrm{MG}}(k,\tau) \hat{D}_{2, \rm{cb}} =& -\BetaFac \mu_{\mathrm{MG}}(k, \tau) D_{1, \rm{cb}}^2(k,\tau)\nonumber\\
&\times \left(1 + \frac{2a^4H^2}{\BetaFac \mu_{\mathrm{MG}}}\gamma_2(k,k/\sqrt{2},k/\sqrt{2}, \tau)\right)\,\comma
\end{align}
which has initial conditions $\hat{D}_{2, \rm{cb}}^{\rm ini} = -\frac{3}{7}$ and
$\frac{d\hat{D}_{2, \rm{cb}}^{\rm ini}}{d\tau} = -\frac{6}{7}\left.\left(\frac{1}{a}\frac{da}{d\tau}\right)\right|_{\tau=\tau_{\rm ini}}$.

\subsection{$\Lambda$CDM with massive neutrinos} \label{ssec:2ndorderLCDMmnu}

When including massive neutrinos at second order instead of first order, we no longer have fitting functions available that can add the effect of massive neutrinos to a growth factor calculated for CDM+baryons in cosmologies without massive neutrinos. In the method that follows, we treat the massive neutrinos as entirely linear such that $\delta_{\nu}=\delta^{(1)}_{\nu}$. Thus the only non-linearity comes from the CDM+baryon component $\delta_{\rm{cb}}=\delta^{(1)}_{\rm{cb}} + \delta^{(2)}_{\rm{cb}}$. This idea has been implemented and tested in {\it N}-body simulations \citep{2009JCAP...05..002B} and in Standard Perturbation Theory (SPT)  \cite{Saito:2008bp,Saito:2009ah, Wong}. Reference \cite{Blas} raised the issue that the treatment of massive neutrinos as purely linear causes problems stemming from the violation of momentum conservation. We discuss the impact of this on our work in Appendix \ref{sect:compnu}.

Now that we are including massive neutrinos, the gravitational potential is sourced by both CDM+baryons and the neutrinos, and thus the Fourier space Poisson equation up to second order is
\begin{align}\label{eq:Poisson2ndLCDMmnu}
\mathcal{F}_x \left[ \nabla_x^2 \Phi_N(\vec{x}, \tau) \right](\vec{k}) = \BetaFac  \left( \delta^{(1)}_{\rm{m}} + \delta^{(2)}_{\rm{m}} \right) = 4\pi G a^4 \overline{\rho}_{\rm{m}}  \left( \delta^{(1)}_{\rm{m}} + \delta^{(2)}_{\rm{m}} \right)\,\period
\end{align}
We can expand the expression $\overline{\rho}_{\rm{m}}\left( \delta^{(1)}_{\rm{m}} + \delta^{(2)}_{\rm{m}} \right)$ as
\begin{align} \label{eq:densityexpansion}
\overline{\rho}_{\rm{m}} \left( \delta^{(1)}_{\rm{m}} + \delta^{(2)}_{\rm{m}} \right) 
= \overline{\rho}_{\rm{cb}} \delta^{(1)}_{\rm{cb}} + \overline{\rho}_{\nu} \delta^{(1)}_{\nu} + \overline{\rho}_{\rm{cb}} \delta^{(2)}_{\rm{cb}}
&= \left( \frac{\overline{\rho}_{\rm{cb}} }{\overline{\rho}_{\rm{m}}} + \frac{\overline{\rho}_{\nu}}{\overline{\rho}_{\rm{m}}} \frac{\delta^{(1)}_{\nu}}{\delta^{(1)}_{\rm{cb}}}  \right) \overline{\rho}_{\rm{m}} \delta^{(1)}_{\rm{cb}} + f_{\mathrm{cb}} {\overline{\rho}_{\rm{m}}} {\delta^{(2)}_{\rm{cb}}}
\nonumber\\ &= \left( f_{\rm{cb}} + f_{\nu} \frac{D_{1, \nu}(k, \tau)}{D_{1, \rm{cb}}(k, \tau)}  \right) \overline{\rho}_{\rm{m}} \delta^{(1)}_{\rm{cb}} + f_{\mathrm{cb}} {\overline{\rho}_{\rm{m}}} {\delta^{(2)}_{\rm{cb}}}
\nonumber\\ &= {\overline{\rho}_{\rm{m}}} \left( \mu_{m_{\nu}}(k, \tau) {\delta^{(1)}_{\rm{cb}}} + f_{\rm cb} {\delta^{(2)}_{\rm{cb}}} \right)\,\comma
\end{align}
such that Eq.~(\ref{eq:Poisson2ndLCDMmnu}) becomes
\begin{align}\label{eq:Poisson2ndLCDMmnu2}
\mathcal{F}_x \left[ \nabla_x^2 \Phi_N(\vec{x}, \tau) \right](\vec{k}) = \BetaFac \left( \mu_{m_{\nu}}(k, \tau) {\delta^{(1)}_{\rm{cb}}} + f_{\rm cb} {\delta^{(2)}_{\rm{cb}}} \right)\,\period
\end{align}
Therefore the Fourier transform with respect to $q$ instead of $x$ is
\begin{multline}
\mathcal{F}_q \left[ {\nabla_x}^2 {\Phi}(\vec{x}, \tau) \right] = \BetaFac \mu_{m_{\nu}}(k, \tau) \tilde{\delta}^{(1)}_{\rm{cb}}(\vec{k}, \tau) + \BetaFac f_{\rm cb} \tilde{\delta}^{(2)}_{\rm{cb}}(\vec{k}, \tau) 
\\+ \BetaFac \int \frac{d^3\vec{k}_1d^3\vec{k}_2}{(2\pi)^3} \delta_D(\vec{k}-\vec{k_{12}}) \left[ \mu_{m_{\nu}}(k, \tau) - \mu_{m_{\nu}}(k_1, \tau) \right]
\\ \times \frac{\vec{k}_1 \cdot \vec{k}_2}{k_2^2} \delta^{(1)}_{\rm{cb}}(\vec{k}_1, \tau) \delta^{(1)}_{\rm{cb}}(\vec{k}_2, \tau) \period
\end{multline}
Inserting this into the full equation of motion up to second order yields
\begin{multline}
k^2 \frac{d^2}{d\tau^2} \left( \phi^{(1)}_{\mathrm{cb}}(k, \tau) + \phi^{(2)}_{\mathrm{cb}}(k, \tau) \right) + \mathcal{F}_q\left[\Psi^{(1)}_{\mathrm{cb}\ j, i}\frac{d^2\Psi^{(1)}_{\mathrm{cb}\ i, j}}{d\tau^2}\right]
\\ = \mathcal{F}_q \left[ {\nabla_x}^2 {\Phi}(\vec{x}, \tau) \right] = \BetaFac \mu_{m_{\nu}}(k, \tau) \tilde{\delta}^{(1)}_{\rm{cb}}(\vec{k}, \tau) + \BetaFac f_{\rm cb} \tilde{\delta}^{(2)}_{\rm{cb}}(\vec{k}, \tau) 
\\+ \BetaFac \int \frac{d^3\vec{k}_1d^3\vec{k}_2}{(2\pi)^3} \delta_D(\vec{k}-\vec{k_{12}}) \left[ \mu_{m_{\nu}}(k, \tau) - \mu_{m_{\nu}}(k_1, \tau) \right] 
\\ \times \frac{\vec{k}_1 \cdot \vec{k}_2}{k_2^2} \delta^{(1)}_{\rm{cb}}(\vec{k}_1, \tau) \delta^{(1)}_{\rm{cb}}(\vec{k}_2, \tau) \,\period
\end{multline}
We can separate out the first and second order terms into two equations. The first order equation is
\begin{align}
k^2 \frac{d^2 \phi^{(1)}_{\mathrm{cb}}(\vec{k}, \tau)}{d\tau^2} = \BetaFac \mu_{m_{\nu}}(k, \tau) \tilde{\delta}^{(1)}_{\rm{cb}}(\vec{k},\tau)\,\period
\end{align}
We use the Fourier space versions of Eq.~(\ref{eq:deltapsi1}) to write
\begin{align} \label{eq:2LPTLCDMmnuinter1}
\left( \frac{d^2}{d\tau^2} - \BetaFac \mu_{m_{\nu}}(k, \tau) \right) {\phi^{(1)}_{\rm{cb}}}(k, \tau) = 0\,\period
\end{align}
The first order equation can be solved numerically at each $(k, \tau)$, where the initial conditions are given by inserting the EdS initial conditions into the fitting formula Eq.~(\ref{eq:D1cb}). The second order equation is
\begin{multline}
k^2 \frac{d^2 \phi^{(2)}_{\mathrm{cb}}(\vec{k}, \tau)}{d\tau^2} + \mathcal{F}_q \left[\Psi^{(1)}_{\mathrm{cb}\ j, i}\frac{d^2\Psi^{(1)}_{\mathrm{cb}\ i, j}}{d\tau^2}\right] = \BetaFac f_{\rm cb} \tilde{\delta}^{(2)}_{\rm{cb}}(\vec{k}, \tau) 
\\+ \BetaFac \int \frac{d^3\vec{k}_1d^3\vec{k}_2}{(2\pi)^3} \delta_D(\vec{k}-\vec{k_{12}}) \left[ \mu_{m_{\nu}}(k, \tau) - \mu_{m_{\nu}}(k_1, \tau) \right] 
\\ \times \frac{\vec{k}_1 \cdot \vec{k}_2}{k_2^2} \delta^{(1)}_{\rm{cb}}(\vec{k}_1, \tau) \delta^{(1)}_{\rm{cb}}(\vec{k}_2, \tau) \,\period
\end{multline}
We use the Fourier space versions of Eq.~(\ref{eq:deltapsi2}) to write
\begin{multline}\label{eq:2LPTLCDMmnuinter12}
k^2 \left( \frac{d^2}{d\tau^2} - \BetaFac f_{\rm cb} \right)\phi^{(2)}_{\mathrm{cb}}(\vec{k}, \tau) = \frac{1}{2} \BetaFac f_{\rm cb} \mathcal{F}\left[\Psi^{(1)}_{\mathrm{cb}\ i, i} \Psi^{(1)}_{\mathrm{cb}\ j, j} + \Psi^{(1)}_{\mathrm{cb}\ j, i} \Psi^{(1)}_{\mathrm{cb}\ i, j} \right] 
\\- \mathcal{F}_q \left[\Psi^{(1)}_{\mathrm{cb}\ j, i}\frac{d^2\Psi^{(1)}_{\mathrm{cb}\ i, j}}{d\tau^2}\right] + \BetaFac \int \frac{d^3\vec{k}_1d^3\vec{k}_2}{(2\pi)^3} \delta_D(\vec{k}-\vec{k_{12}}) \left[ \mu_{m_{\nu}}(k, \tau) - \mu_{m_{\nu}}(k_1, \tau) \right]
\\ \times \frac{\vec{k}_1 \cdot \vec{k}_2}{k_2^2} \delta^{(1)}_{\rm{cb}}(\vec{k}_1, \tau) \delta^{(1)}_{\rm{cb}}(\vec{k}_2, \tau) \,\period
\end{multline}
If we define $\phi^{(2)}_{\rm cb}(\vec{k}, \tau)$ as in Eq.~(\ref{eq:psi2lptintegral}), rewrite the final term in Eq.~(\ref{eq:2LPTLCDMmnuinter1}) using the first order solution, and write out the Fourier transforms explicitly, then the second order equation becomes
\begin{multline}\label{eq:2LPTLCDMmnufull}
\left( \frac{d^2}{d\tau^2} - \BetaFac f_{\rm cb} \right) D_{2, \mathrm{cb}}(\vec{k}, \vec{k_1}, \vec{k_2}, \tau) 
=  \BetaFac D_{1, \mathrm{cb}}(k_1, \tau) D_{1, \mathrm{cb}}(k_2, \tau) 
\\ \times \left\{ \left[ 2 \mu_{m_{\nu}}(k_1, \tau) - f_{\rm cb} \right] \frac{(\vec{k_1}.\vec{k_2})^2}{k_1^2 k_2^2}  \right. 
\\ \left.- f_{\rm cb} + \left[ \mu_{m_{\nu}}(k, \tau) - \mu_{m_{\nu}}(k_1, \tau) \right] \frac{\vec{k}_1 \cdot \vec{k}_2}{k_2^2} \right\} \,\period
\end{multline}

As for the case of modified gravity without massive neutrinos, the integral in Eq.~(\ref{eq:psi2lptintegral}) would be very slow to evaluate at each time-step. To circumvent this problem, we make an approximation by redefining $\phi^{(2)}_{\rm cb}$ as
\begin{align}\label{eq:phi2approx2}
\phi^{(2)}_{\rm{cb}}(\vec{k},\tau) =& -\frac{\hat{D}_{2, \rm{cb}}(k,\tau)}{2k^2}&\nonumber \\ &\times \int\frac{{\rm d}^3\vec{k}_1{\rm d}^3\vec{k}_2}{(2\pi)^3}\delta_D(\vec{k}-\vec{k}_{12}) \delta^{(1)}_{\rm{cb}}(\vec{k_1},\tau_{\rm ini})\delta^{(1)}_{\rm{cb}}(\vec{k_2},\tau_{\rm ini})\left(1 - \frac{(\vec{k_1}\cdot\vec{k_2})^2}{k_1^2k_2^2}\right)\,\comma
\end{align}
and setting $\mu_{m_{\nu}} \rightarrow f_{\rm cb}$, which leads to the following equation for $\hat{D}_{2, \rm{cb}}(k,\tau)$:
\begin{align}\label{eq:2LPTLCDMmnuapprox2}
\left( \frac{d^2}{d\tau^2} - \BetaFac f_{\rm cb} \right) \hat{D}_{2, \mathrm{cb}}(k, \tau) 
= - \BetaFac f_{\rm cb} D^2_{1, \mathrm{cb}}(k, \tau)\,\period
\end{align}
For $f_{\rm cb} = 1$ we recover the equation for $\Lambda$CDM without massive neutrinos. For a matter dominated Universe and for scales smaller than the neutrino free-streaming scale we have $D_{2,\rm cb} \simeq - \frac{3f_{\rm cb}}{3f_{\rm cb} + 4(1-p_{\rm cb})^2}D_{1,\rm cb}^2$ which can be used to set the initial conditions when solving it numerically. The $\mu_{m_{\nu}} \rightarrow f_{\rm cb}$ approximation was previously made in \cite{Saito:2008bp,Saito:2009ah} for a $\Lambda$CDM+massive neutrino cosmology in Standard Perturbation Theory. They argued that the small value of $f_{\nu}$ suppresses the non-linear corrections to the above approximation of treating the massive neutrinos as an exclusively linear density perturbation. We have tested the effect of using $\mu_{m_\nu}(k)$ instead of $f_{cb}$ in Eq.~(\ref{eq:2LPTLCDMmnuapprox2}) for the second order growth-factor in our COLA implementation (to be presented in the upcoming section). This change was found to have an negligible effect ($\lesssim 0.1-0.5$ \% ) on the total matter power-spectrum for wavenumbers $k \lesssim 1\ h/$Mpc.

\begin{figure*}
\includegraphics[width=0.9\textwidth]{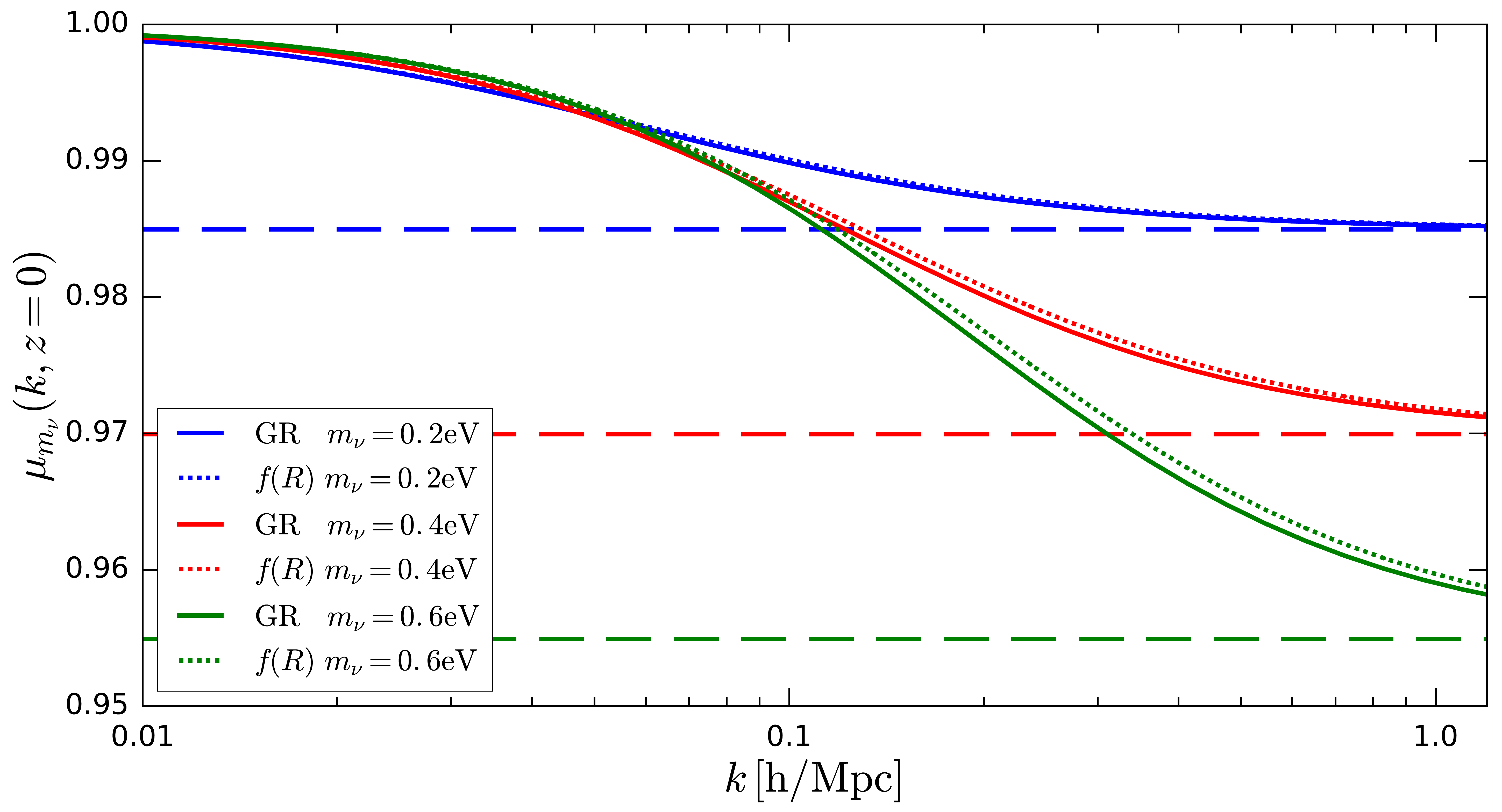}
\caption{The value of $\mu_{m_{\nu}}$ as a function of $k$ calculated using Eq.~(\ref{eq:densityexpansion}) at $z=0$ for both GR and F4 cosmologies with neutrinos of mass $m_{\nu}=[0.2, 0.4, 0.6]$ eV. The horizontal dashed lines plotted are the values of $f_{\rm cb}$ at each neutrino mass, which highlight the consequences of setting $\mu_{m_{\nu}} \rightarrow f_{\rm cb}$ in the approximate second order growth factor equations Eqs.~(\ref{eq:2LPTLCDMmnuapprox2}) and (\ref{eq:2LPTMGmnuapprox}).}
\label{fig:mu_mnu}
\end{figure*}

In Figure~\ref{fig:mu_mnu} we show the difference between $\mu_{m_{\nu}}$, calculated using Eq.~(\ref{eq:densityexpansion}), and $f_{\rm cb}$ in the range $0.01 \leq k \leq 1.0\ h/\mathrm{Mpc}$ to highlight the consequences of making this approximation. Figure~\ref{fig:mu_mnu} shows that the approximation is less important as $k \rightarrow 1.0\ h/\mathrm{Mpc}$, but also shows that the approximation becomes less accurate as $m_{\nu}$ increases. Specifically, at $k=0.01\ h/\mathrm{Mpc}$, the percentage difference between $\mu_{m_{\nu}}$ and $f_{\rm cb}$ is 1.5/3.0/4.5$\%$ for $m_{\nu}=0.2/0.4/0.6$ eV respectively.

\subsection{Scale-dependent modified gravity with massive neutrinos} \label{ssec:2ndorderMGmnu}

In the case with both modified gravity and massive neutrinos, the Fourier space Poisson equation up to second order, with the potential $\Phi$ being sourced by both CDM+baryons and massive neutrinos, is
\begin{align}\label{eq:2LPTMGandmu}
\mathcal{F}_x \left[ \nabla_x^2 \Phi \right] = &\BetaFac \mu_{\mathrm{MG}}(k,\tau) \left( \delta^{(1)}_{\rm{m}} + \delta^{(2)}_{\rm{m}} \right) \nonumber\\ &+ a^4H^2\int\frac{{\rm d}^3\vec{k}_1{\rm d}^3\vec{k}_2}{(2\pi)^3}\delta^{(1)}_{\rm{cb}}(\vec{k_1}, \tau)\delta^{(1)}_{\rm{cb}}(\vec{k_2}, \tau)\gamma_2^{\rm E}(\vec{k},\vec{k_1},\vec{k_2}, \tau)\,\period
\end{align}
Repeating the working of Sections \ref{ssec:2ndorderMG} and \ref{ssec:2ndorderLCDMmnu}, leads to
\begin{multline}\label{eq:Poisson2ndMGmu2}
\mathcal{F}_q \left[ {\nabla_x}^2 {\Phi}(\vec{x}, \tau) \right] = \BetaFac \mu_{\mathrm{MG}}(k,\tau)\mu_{m_{\nu}}(k,\tau) \tilde{\delta}^{(1)}_{\rm{cb}}(\vec{k}, \tau) + \BetaFac f_{\rm cb}\mu_{\rm MG}(k, \tau) \tilde{\delta}^{(2)}_{\rm{cb}}(\vec{k}, \tau)
\\+ a^4 H^2\int \frac{d^3\vec{k}_1d^3\vec{k}_2}{(2\pi)^3} \delta_D(\vec{k}-\vec{k_{12}})\gamma_2(\vec{k},\vec{k_1},\vec{k_2}, \tau) \delta^{(1)}_{\rm{cb}}(\vec{k_1}, \tau) \delta^{(1)}_{\rm{cb}}(\vec{k_2}, \tau) \comma
\end{multline}
where again we define $\gamma_2 = \gamma_2^{\rm E} + \frac{3}{2} \Omega_{\rm m}(\tau) \left[ \mu_{\mathrm{MG}}(k,\tau)\mu_{m_{\nu}}(k,\tau) - \mu_{\mathrm{MG}}(k_1,\tau)\mu_{m_{\nu}}(k_1,\tau) \right] \frac{\vec{k}_1 \cdot \vec{k}_2}{k_2^2}$. Inserting this expression into the equation of motion yields the first order equation
\begin{align}
\left( \frac{d^2}{d\tau^2} - \BetaFac \mu_{\rm MG}(k, \tau) \mu_{m_{\nu}}(k, \tau) \right) \phi^{(1)}_{\rm cb} (k, \tau) = 0\,\comma
\end{align}
which can be solved numerically at each $(k, \tau)$, where the initial conditions are given by inserting the EdS initial conditions into the fitting formula Eq.~(\ref{eq:D1cb}). The second order equation is
\begin{multline}\label{eq:2LPTMGmnufull}
\left[ \frac{d^2}{d\tau^2} - \BetaFac f_{\rm cb}\mu_{\rm MG}(k, \tau) \right] D_{2, \rm{cb}}(\vec{k},\vec{k_1},\vec{k_2},\tau) 
\\ = \BetaFac \left\{  \left[ 2 \mu_{\rm MG}(k_1, \tau) \mu_{m_{\nu}}(k_1, \tau) - f_{\rm cb}\mu_{\rm MG}(k, \tau) \right]\frac{ ( \vec{k}_1 \cdot \vec{k}_2 )^2}{k_1^2 k_2^2} - f_{\rm cb}\mu_{\rm MG}(k, \tau) \right.
\\ \left. - \frac{2a^4 H^2 \gamma_2(\vec{k}, \vec{k_1}, \vec{k_2}, \tau)}{\BetaFac} \right\} D_{1, \mathrm{cb}}(k_1, \tau) D_{1, \mathrm{cb}}(k_2, \tau) .
\end{multline}
To speed up the calculation, we again make an approximation by redefining $\phi^{(2)}_{\rm cb}$ as
\begin{align}\label{eq:phi2approx3}
\phi^{(2)}_{\rm{cb}}(\vec{k},\tau) =& -\frac{\hat{D}_{2, \rm{cb}}(k,\tau)}{2k^2}&\nonumber \\ &\times \int\frac{{\rm d}^3\vec{k}_1{\rm d}^3\vec{k}_2}{(2\pi)^3}\delta_D(\vec{k}-\vec{k}_{12}) \delta^{(1)}_{\rm{cb}}(\vec{k_1},\tau_{\rm ini})\delta^{(1)}_{\rm{cb}}(\vec{k_2},\tau_{\rm ini})\left(1 - \frac{(\vec{k_1}\cdot\vec{k_2})^2}{k_1^2k_2^2}\right)\,\comma
\end{align}
and setting $\mu_{m_{\nu}} \rightarrow f_{\rm cb}$, which leads to the following equation for $\hat{D}_{2, \mathrm{cb}}$:
\begin{align}\label{eq:2LPTMGmnuapprox}
\left( \frac{d^2}{d\tau^2} -\BetaFac \mu_{\rm MG}(k, \tau) f_{\rm cb} \right) \hat{D}_{2, \mathrm{cb}}(k, \tau)
= &- \BetaFac \mu_{\rm MG}(k, \tau) f_{\rm cb} D^2_{1, \mathrm{cb}}(k, \tau) 
\nonumber \\ &\times \left(1 + \frac{2a^4H^2}{\BetaFac \mu_{\mathrm{MG}} f_{\rm cb}}\gamma_2(k,k/\sqrt{2},k/\sqrt{2}, \tau)\right)\,\period
\end{align}
For modified gravity models which only have late-time effects we can use the same initial conditions as for $\Lambda$CDM discussed above. Figure~\ref{fig:mu_mnu} shows that there is a negligible difference between the F4 model of $f(R)$ gravity and GR in the comparison between the values of $\mu_{m_{\nu}}$ and $f_{\rm cb}$ in the range $0.01 \leq k \leq 1.0\ h/\mathrm{Mpc}$.

\section{COLA Implementation} \label{sec:COLA}

The implementation of massive neutrinos in the particle mesh part of the COLA algorithm is the grid-based method suggested in \cite{2009JCAP...05..002B}. This method has been demonstrated to produce a matter power-spectrum that is accurate to $< 1\%$ for neutrino masses $\sum m_\nu \lesssim 0.6$ eV. When we create the initial conditions for the CDM (CDM+baryon) particles we use the same initial seed to create a realisation of massive neutrinos using
\begin{align}
\delta_\nu(\vec{k},\tau_{\rm ini}) = \delta_{\rm cb}(\vec{k},\tau_{\rm ini}) \frac{T_\nu(k,\tau_{\rm ini})}{T_{\rm cb}(k,\tau_{\rm ini})} = \delta_{\rm cb}(\vec{k},\tau_{\rm ini}) \frac{D_{1, \nu}(k,\tau_{\rm ini})}{D_{1, \rm cb}(k,\tau_{\rm ini})}\,\comma
\end{align}
where $T_\nu$ and $T_{\rm cb}$ are the transfer functions of massive neutrinos and CDM+baryons respectively. 

The massive neutrinos are kept in Fourier space for the duration of the simulation and are added to the source of the Poisson equation
\begin{align}\label{eq:nlpoissoneq}
-k^2\Phi(\vec{k},\tau) = \frac{3}{2}\Omega_{\rm{m}} a\left[f_{cb}\delta_{\rm cb}(\vec{k},\tau) + f_\nu\delta_\nu(\vec{k},\tau)\right]\,\comma
\end{align}
where the neutrino density at a given time $\tau$ is computed as
\begin{align}\label{eq:neutrinodensitypoisson}
\delta_\nu(\vec{k},\tau) = \delta_\nu(\vec{k},\tau_{\rm ini}) \frac{T_\nu(k,\tau)}{T_\nu(k,\tau_{\rm ini})} = \delta_\nu(\vec{k},\tau_{\rm ini}) \frac{D_{1, \nu}(k,\tau)}{D_{1, \nu}(k,\tau_{\rm ini})}\,\period
\end{align}
In Appendix~\ref{sect:compnu} we show a comparison of this scheme to an alternative scheme of modeling the non-linear neutrino density.

As for the COLA specific part we use the (scale-dependent) growth factors discussed in the previous sections. We have compared computing the growth factors using the fitting functions to directly solving the growth-ODEs with $\mu_\nu(k,\tau) = f_{cb} + f_\nu\frac{T_\nu(k,\tau)}{T_{cb}(k,\tau)}$ computed using transfer functions from {\tt{CAMB}} or its alternatives. The difference between these two approaches was found to be negligible.

As long as the cosmological model we simulate already has scale-dependent growth then the additional computational cost of adding in neutrinos this way is almost negligible, but it does require some extra memory as we need to store the initial neutrino density field.

\section{Simulation Results} \label{sec:Results}

\begin{figure*}
\includegraphics[width=0.9\textwidth]{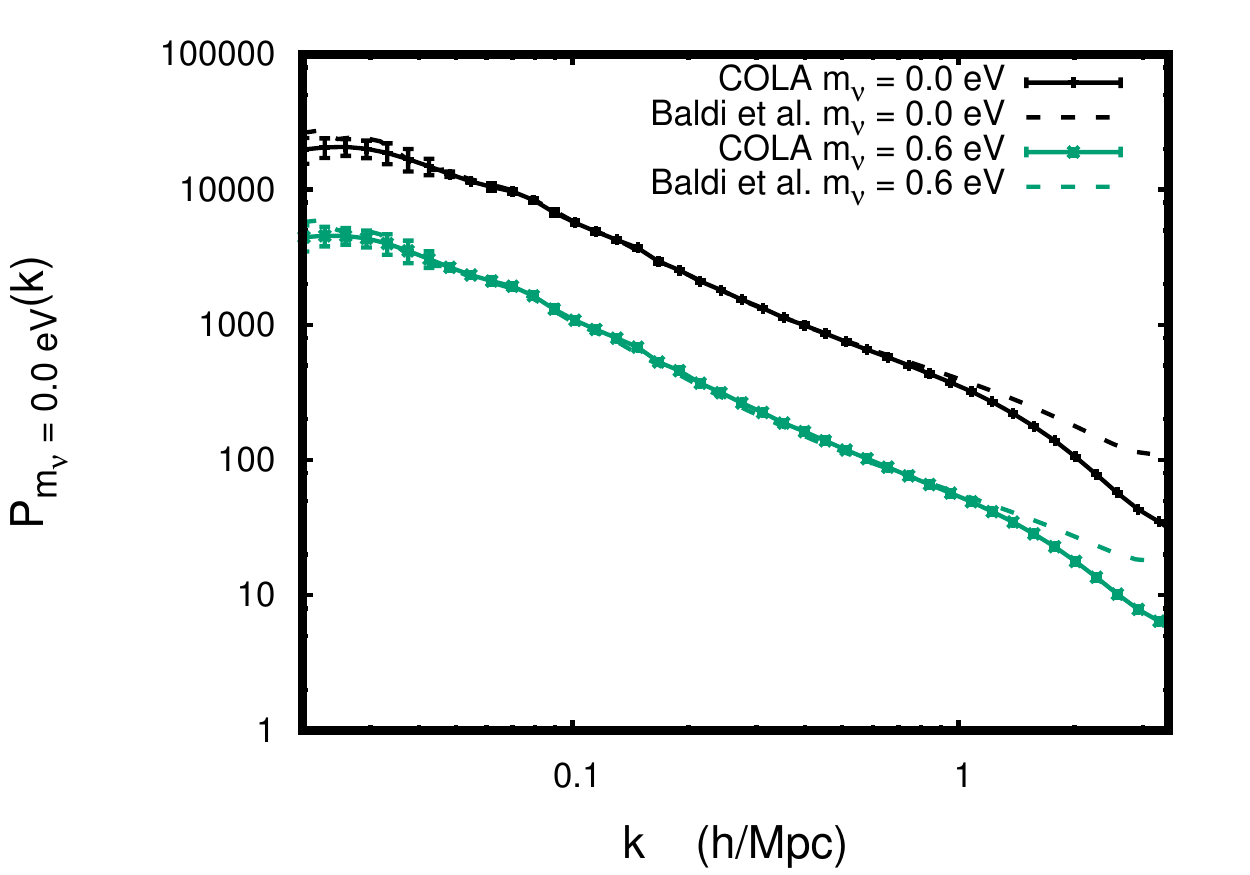}
\caption{The CDM matter power-spectrum $P(k,z=0)$ for $\Lambda$CDM in our COLA simulations compared with the result of \cite{Baldi2014}. The $m_\nu  = 0.6$ eV results are offset by a factor of $0.25$.}
\label{fig:pofk_full_lcdm_z0}
\end{figure*}

We ran 5 COLA {\it N}-body simulations in a box of $B = 512$ Mpc$/h$ with $N = 512^3$ particles. The simulations were performed using the \texttt{MG-PICOLA} code. A smaller box of $B = 256$ Mpc$/h$ with the same number of particles was used to check the convergence of the results, and Figure~\ref{fig:pofk_full_lcdm_z0} shows, through comparison to the full {\it N}-body simulations of \cite{Baldi2014}, that the CDM matter power-spectrum in our simulations can be trusted to percent level up to $k\sim 0.5-0.7~h/$Mpc. As found in our previous paper, the relative enhancement of the power-spectrum (i.e. when considering ratios of power-spectra as shown in the figures below) is accurate to larger $k$ values. The cosmological parameters for the simulations can be found in Table~\ref{tab:param} and these are the same parameters as used by \cite{Baldi2014} where they performed combined massive neutrino and modified gravity simulations using a modified version of the simulation code \texttt{Gadget} \citep{2005MNRAS.364.1105S,2013MNRAS.436..348P}. We will use these simulations to compare our results below. Ideally we would like to have run our simulations using exactly the same initial seed as the {\it N}-body simulations, however this was not available at the time we wrote this paper and we leave such a detailed comparison to future work.

\begin{table}
\begin{center}
\begin{tabular}{ |c|c|c|c| }
 \hline
$m_\nu$ (eV)  & $\Omega_{\rm CDM}$ & $\Omega_\nu$ & $\sigma_8$ ($\Lambda\rm CDM$)\\
 \hline
 0.0 & 0.2685 & 0.0    & 0.850\\
 0.2 & 0.2637 & 0.0048 & 0.798\\
 0.4 & 0.259  & 0.0095 & 0.752\\
 0.6 & 0.2542 & 0.0143 & 0.712\\
 \hline
\end{tabular}
\caption{The cosmological parameters for the simulations performed in this paper. Common to all simulations are $\Omega_{\rm m} = 0.3175$, $\Omega_b = 0.049$, $n_s = 0.966$, $A_s=2.215\cdot 10^{-9}$ and $h = 0.671$.}
\label{tab:param}
\end{center}
\end{table}

\begin{figure*}
\includegraphics[width=0.9\textwidth]{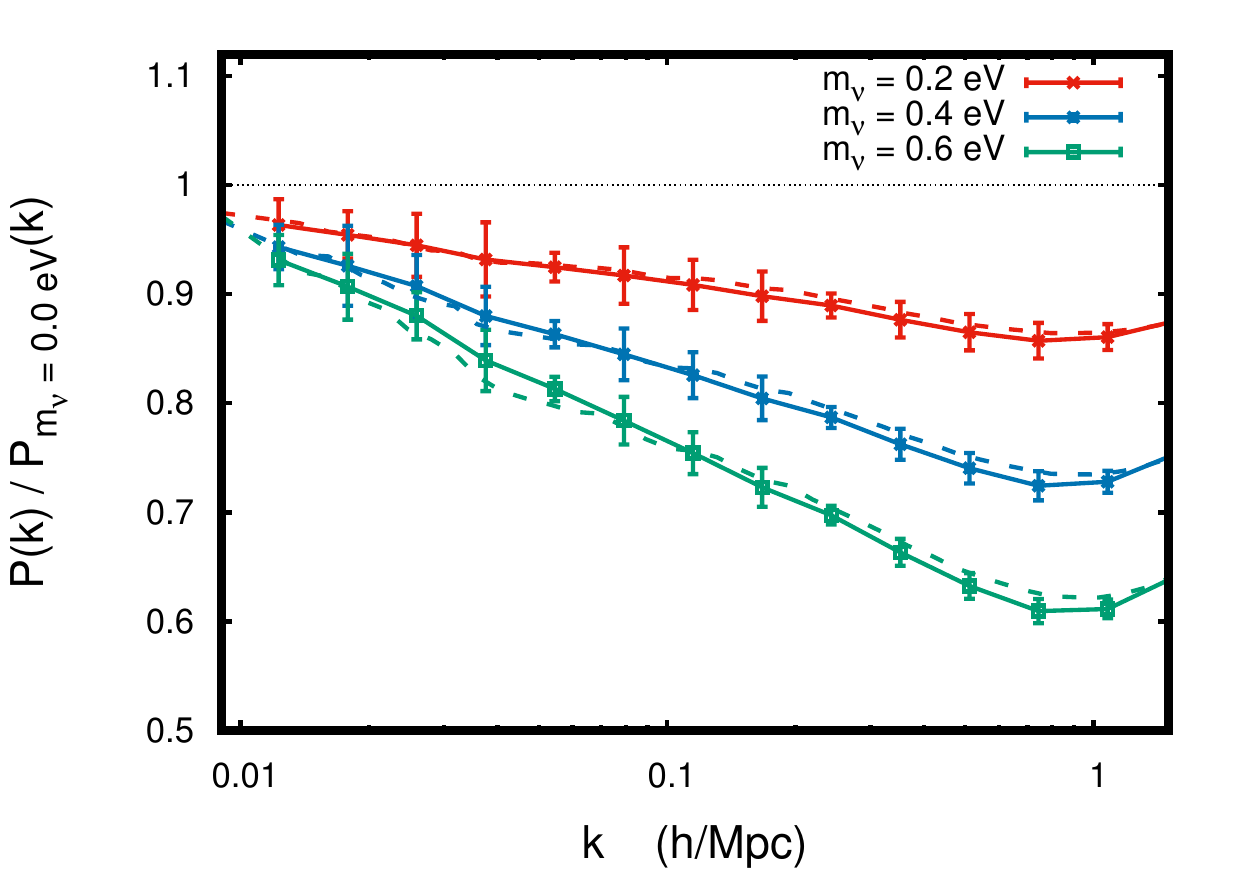}
\includegraphics[width=0.9\textwidth]{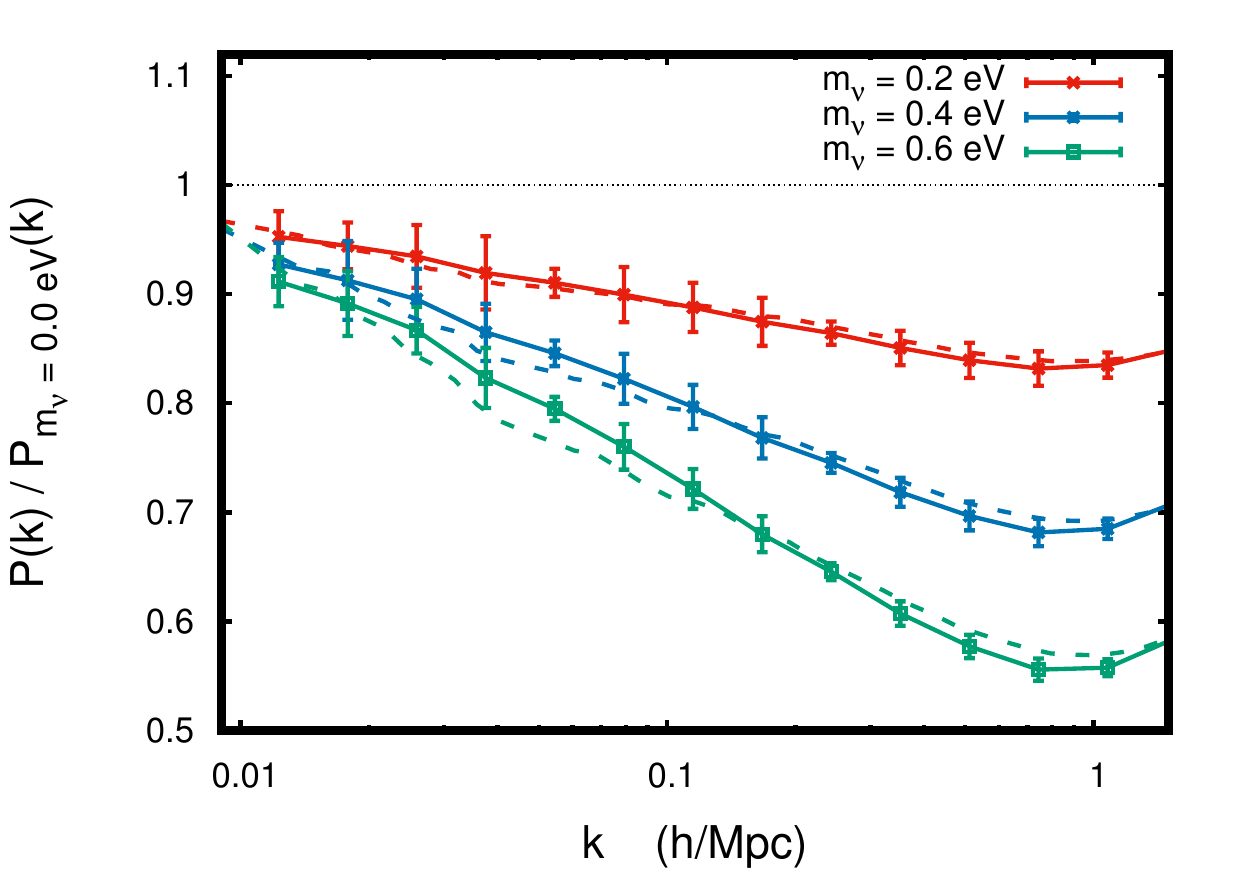}
\caption{The matter power-spectrum $P(k,z=0)$ for several values of the sum of neutrino masses relative to the power-spectrum with $m_\nu = 0.0$ for $\Lambda$CDM. The solid lines shows the result of \cite{Baldi2014}. The top panel shows the CDM+baryon power-spectrum and the bottom panel shows the total power-spectrum.}
\label{fig:pofk_lcdm_z0}
\end{figure*}

In Figure~\ref{fig:pofk_lcdm_z0} we show the suppression of the power-spectrum in $\Lambda$CDM as a function of neutrino mass in our simulations compared to the {\it N}-body results of \cite{Baldi2014}. We show power-spectrum results for both CDM+baryons $P_{\rm cb}(k,z) = \left<|\delta_{\rm{cb}}(\vec{k},z)|^2\right>$ and the total matter content (CDM+baryons+massive neutrinos) $P_{\rm m}(k,z) = \left<|f_{\rm{cb}}\delta_{\rm{cb}}(\vec{k},z) + f_\nu\delta_\nu(\vec{k},z)|^2\right>$.

\begin{figure*}
\includegraphics[width=0.9\textwidth]{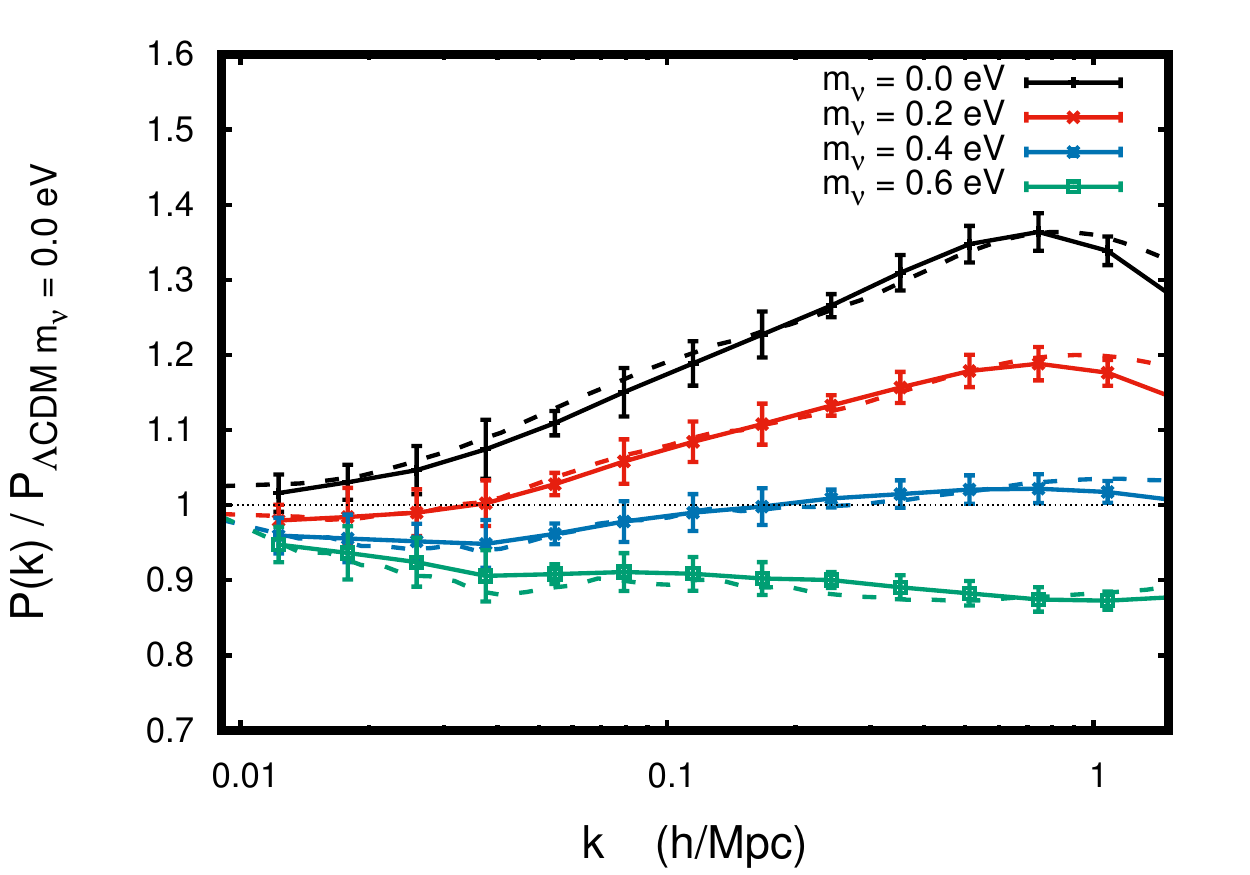}
\includegraphics[width=0.9\textwidth]{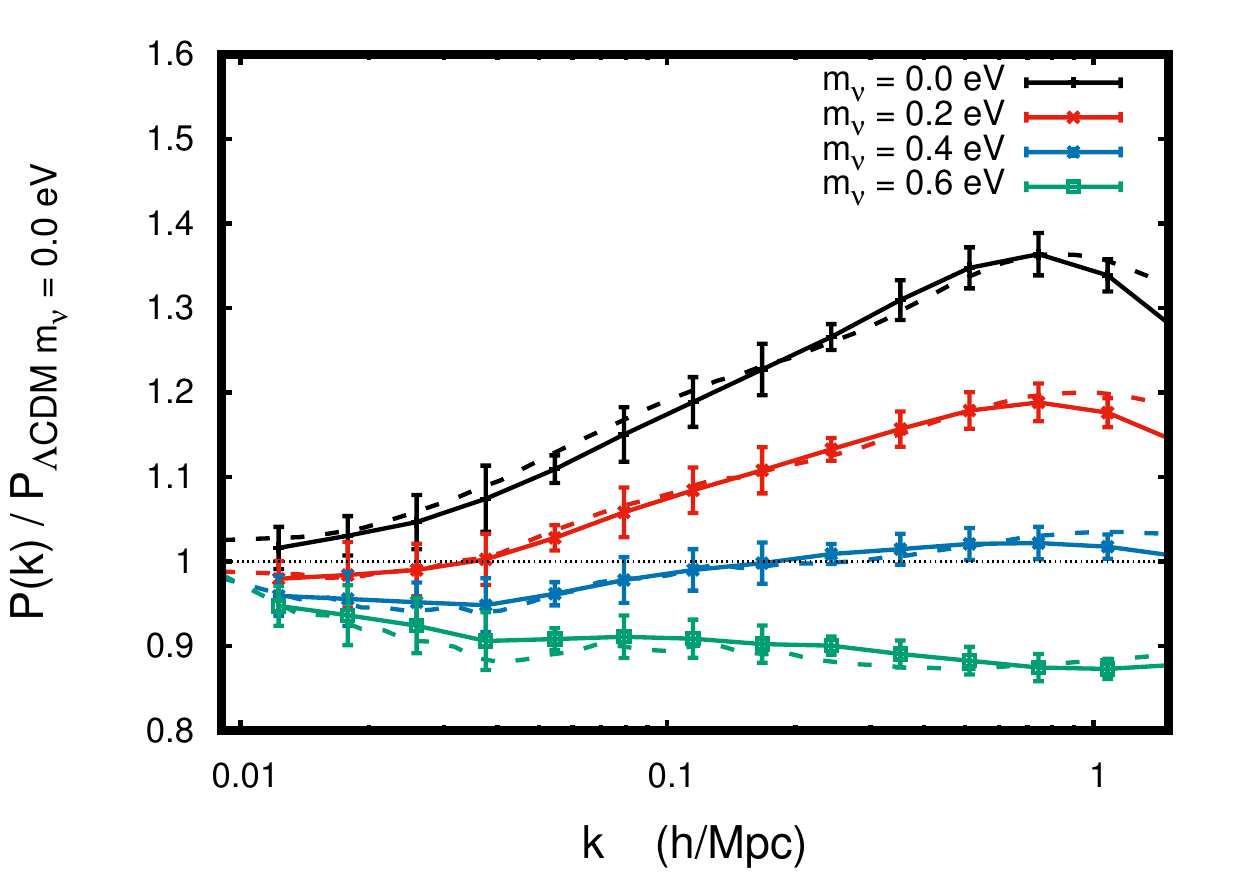}
\caption{The matter power-spectrum $P(k,z=0)$ for several values of the sum of neutrino masses relative to the power-spectrum with $m_\nu = 0.0$ for $f(R)$ gravity with $|f_{R0}| = 10^{-4}$. The solid lines shows the result of \cite{Baldi2014}. The top panel shows the CDM+baryon power-spectrum and the bottom panel shows the total power-spectrum.}
\label{fig:pofk_fofr_z0}
\end{figure*}

In Figure~\ref{fig:pofk_fofr_z0} we show the results from simulations where we have both massive neutrinos and modified gravity. For the particular $f(R)$ model we study here having a total neutrino mass of $m_\nu \sim 0.4$ eV is seen to lead to a power-spectrum very similar to that of a standard $\Lambda$CDM model with massless neutrinos. This illustrates the degeneracy of massive neutrinos (suppressing growth) and modified gravity (enhancing growth).

Our COLA implementation gives power-spectrum (both for CDM and for the total matter) results that agree to $\lesssim 1\%$ accuracy for $k\lesssim 1~h/$Mpc to full {\it N}-body simulations of \cite{Baldi2014} for both $\Lambda$CDM and $f(R)$.

\begin{figure*}
\includegraphics[width=0.9\textwidth]{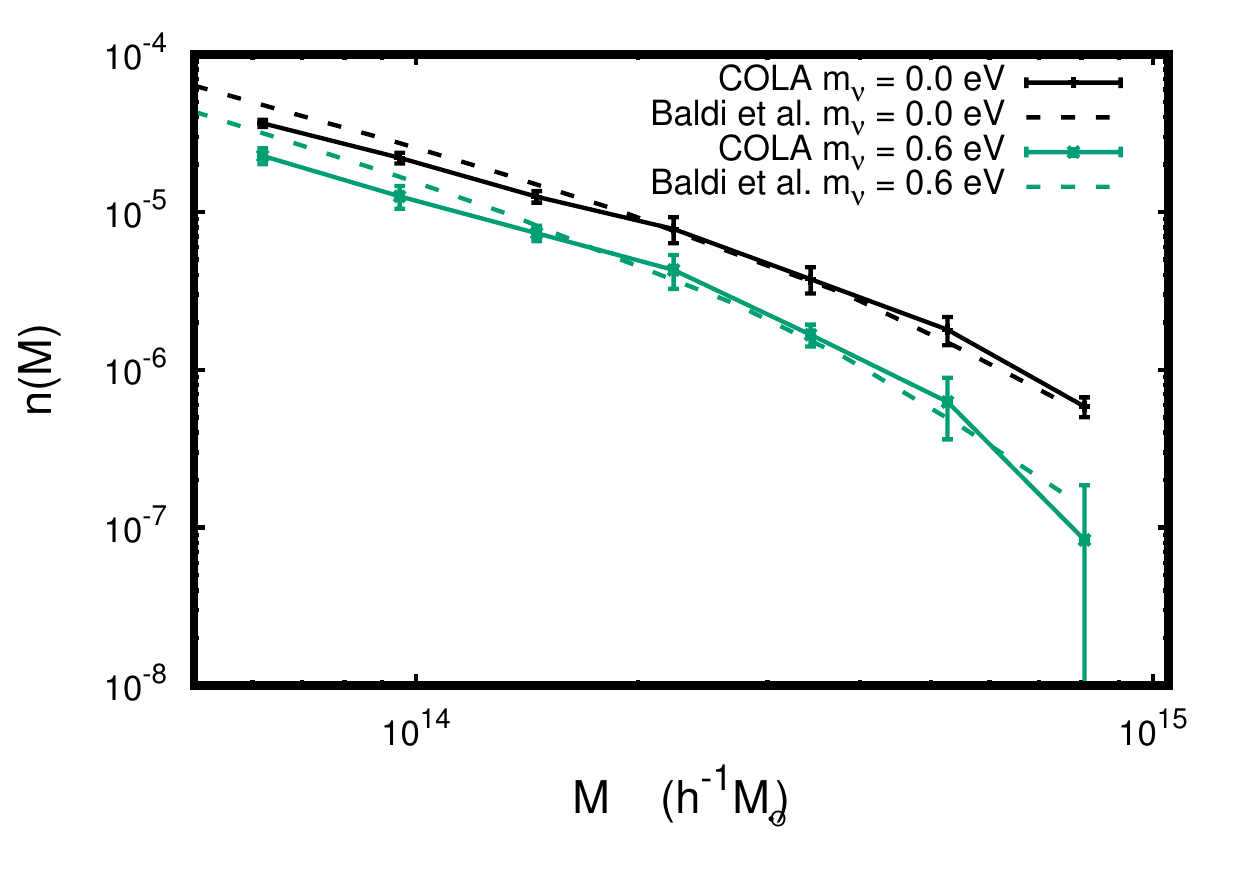}
\includegraphics[width=0.9\textwidth]{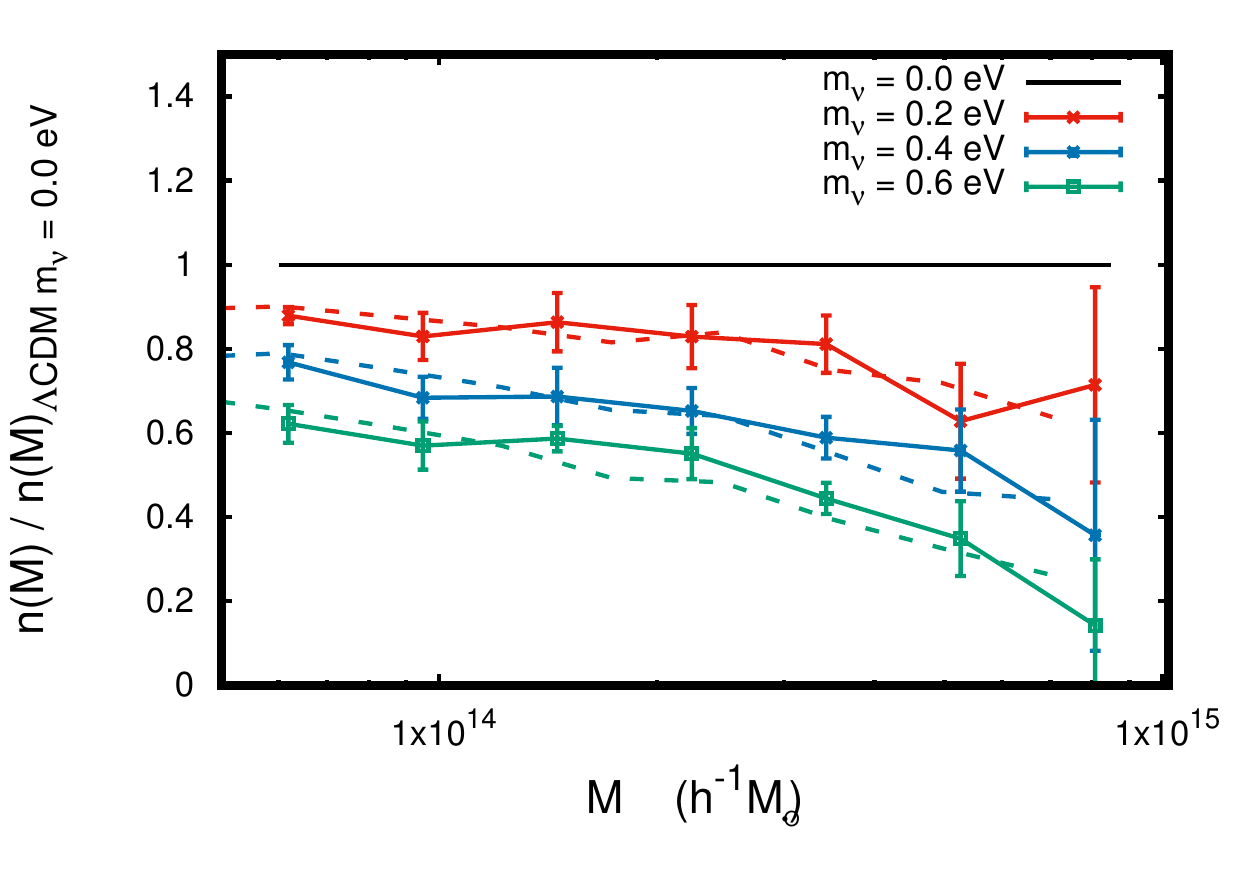}
\caption{The halo mass-function $n(M,z=0)$ for $\Lambda$CDM for several values of the sum of neutrino masses relative to the mass-function with $m_\nu = 0.0$. The dashed lines shows the results from \cite{Baldi2014}.}
\label{fig:nofm_lcdm_z0}
\end{figure*}

\begin{figure*}
\includegraphics[width=0.9\textwidth]{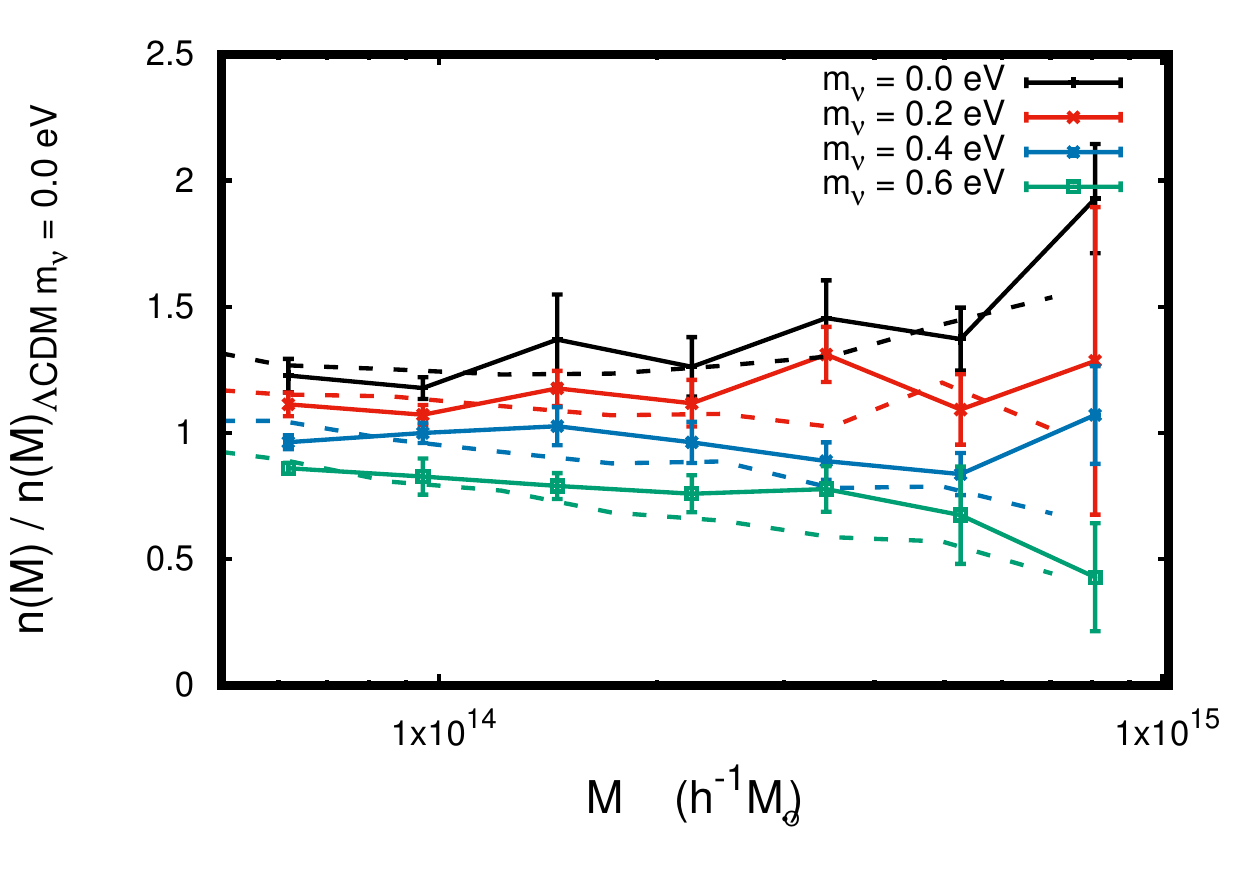}
\caption{The halo mass-function $n(M,z=0)$ for $f(R)$ for several values of the sum of neutrino masses relative to the mass-function in $\Lambda$CDM with $m_\nu = 0.0$. The dashed lines shows the results from \cite{Baldi2014}.}
\label{fig:nofm_fofr_z0}
\end{figure*}

In Figures~\ref{fig:nofm_lcdm_z0} and \ref{fig:nofm_fofr_z0} we show the results for the halo mass function computed using \texttt{Rockstar} \citep{2013ApJ...762..109B}. We note that the results of \cite{Baldi2014} were computed using a different halo-finder (\texttt{SUBFIND}) so the results are not directly comparable; however, the enhancement of the halo mass-function generally shows a good agreement.

\section{Conclusions}\label{sec:Conclusion}

Massive neutrinos, just like many modified gravity theories, lead to scale-dependent growth of matter perturbations. The effect can be described by an effective gravitational constant $\mu_{m_{\nu}}(k,\tau)$. We have implemented massive neutrinos in the COLA {\it N}-body code \texttt{MG-PICOLA} which allows for scale-dependent growth factors. For the particle mesh part of the COLA algorithm we use the grid-based method of \cite{2009JCAP...05..002B} where massive neutrinos are kept in Fourier space and evolved linearly according to the neutrino growth factors. This method has previously been shown to be a good approximation for $\sum m_\nu \lesssim 0.6$ eV. A comparison to full {\it N}-body simulations of massive neutrino cosmologies, both for $\Lambda$CDM and $f(R)$ gravity, shows that we can match {\it N}-body results to percent level accuracy in both the total and CDM matter power-spectra with this approach.

We have also shown that the Eisenstein-Hu fitting formulae for the growth factors in massive neutrino cosmologies are a good approximation for a wide range of modified gravity theories as long as we replace the $\Lambda$CDM growth factor by the modified gravity counterpart.

In order to be able to judge the accuracy of our scheme more directly we would need to do a comparison to full {\it N}-body simulations where we use exactly the same initial conditions in COLA. This was not available to us at the time this study was performed, but such a comparison is something we plan to do in the future. In future work we also intend to study the degeneracy of massive neutrinos and modified gravity in detail and find a way to distinguish this case from $\Lambda$CDM; for example through signatures in redshift-space distortions.

\section*{Acknowledgement}

We would like to thank Marco Baldi for providing us with the data for the $f(R)$ + massive neutrino {\it N}-body simulations. BSW is supported by the U.K. Science and Technology Facilities Council (STFC) research studentship. KK and HAW are supported by the European Research Council through 646702 (CosTesGrav). KK is also supported by the UK Science and Technologies Facilities Council grants ST/N000668/1.

\clearpage
\bibliographystyle{JHEP}
\bibliography{References}

\begin{appendix}

\section{Modified gravity models}\label{sec:MGmodels}

In this section we give a brief overview of the two modified gravity models we are using in this paper focusing on the equations that are needed for our COLA implementation. For a more thorough review of these models, and modified gravity in general, see \cite{CliftonMGReview,2016RPPh...79d6902K}.

\subsection{$f(R)$ gravity}
For $f(R)$ gravity \cite{fofr1} the growth of linear perturbations is determined by
\begin{align}
\mu_{\mathrm{MG}}(k,a) = 1 + \frac{1}{3}\frac{k^2}{k^2 + a^2m^2(a)}\comma
\end{align}
where $m(a)$ depends on the model in question. For the Hu-Sawicki model \cite{fofr2}, which is the $f(R)$ model we will consider in this paper, $m(a)$ is given by
\begin{align}\label{eq:mofahus}
m^2(a) = \frac{1}{3f_{RR}} = \frac{H_0^2(\Omega_{\rm{m}} + 4\Omega_\Lambda)}{(n+1)|f_{R0}|} \left(\frac{\Omega_{\rm{m}} a^{-3} + 4\Omega_\Lambda}{\Omega_{\rm{m}} + 4\Omega_\Lambda}\right)^{n+2}\comma
\end{align}
where 
\begin{align}
f_R(a) = f_{R0}\left(\frac{\Omega_{\rm{m}} + 4\Omega_\Lambda}{\Omega_{\rm{m}}a^{-3} + 4\Omega_\Lambda}\right)^{n+1}\period
\end{align}
The $\gamma_2^E$ factor\footnote{Note that $\gamma_2$ as used in Eq.~(\ref{eq:2LPTMGmnuapprox}) is defined via $\gamma_2(k,k_1,k_2,\tau) = \gamma_2^{\rm E}(k,k_1,k_2,\tau) + \frac{3}{2}\Omega_{\rm m}(\tau)[\mu_{\rm MG}(k,\tau)\mu_\nu(k,\tau) - \mu_{\rm MG}(k_1,\tau)\mu_\nu(k_1,\tau)]\frac{\vec{k}_1\cdot\vec{k}_2}{k_2^2}$.} is likewise given by \cite{BoseKoyama2016}
\begin{align}
\gamma_2^E = -\frac{9\Omega_{\rm{m}}^2}{48a^6|f_{R0}|^2}\left(\frac{k}{aH}\right)^2 \times \frac{(\Omega_{\rm{m}} a^{-3} + 4\Omega_\Lambda)^5}{(\Omega_{\rm{m}} +4\Omega_\Lambda)^4}\frac{1}{\Pi(k,a)\Pi(k_1,a)\Pi(k_2,a)}\comma
\end{align}
where
\begin{align}
\Pi(k,a) = \left(\frac{k}{aH_0}\right)^2 + \frac{(\Omega_{\rm{m}}a^{-3} + 4\Omega_\Lambda)^3}{2|f_{R0}|(\Omega_{\rm{m}} + 4\Omega_\Lambda)^2}\period
\end{align}

\subsection{Symmetron}

In the symmetron model \citep{2010PhRvL.104w1301H}:
\begin{align}
\mu_{\mathrm{MG}}(k, a) = 1 + \frac{2 \beta^2(a) k^2}{k^2 + a^2m^2(a)}\,\comma
\end{align}
\begin{align}
\gamma_2^{\rm E}(k, a) &= \frac{m^2(a)\frac{dm^2(a)}{da}\beta^2(a) \Omega_m}{2H_0^4\Pi(k,a)\Pi(k_1,a)\Pi(k_2,a)}\frac{k^2}{a^4H^2}\nonumber\\
&= \frac{3\Omega_m\beta_\star^2}{2\xi_\star^4}\frac{a_\star^3 k^2}{a^8H^2\Pi(k,a)\Pi(k_1,a)\Pi(k_2,a)}\left(1 - \frac{a_\star^3}{a^3}\right)^2\,\comma
\end{align}
if $a>a_\star$ and $0$ otherwise where
\begin{align}
\beta(a) = \begin{cases}
\beta_{\star} \sqrt[]{1-\frac{a_{\star}^3}{a^3}} & \mathrm{if}\ a>a_{\star} \\ 0 & \mathrm{otherwise}
\end{cases}\,\comma
\end{align}
\begin{align}
m(a) = \begin{cases}
\frac{H_0}{\xi_{\star}} \sqrt[]{1-\frac{a_{\star}^3}{a^3}} & \mathrm{if}\ a>a_{\star} \\ 0 & \mathrm{otherwise}
\end{cases}\,\period
\end{align}

\subsection{Dilaton}

In the dilaton model \citep{2010PhRvD..82f3519B}:
\begin{align}
\mu_{\mathrm{MG}}(k, a) = 1 + \frac{2 \beta^2(a) k^2}{k^2 + a^2m^2(a)}\,\comma
\end{align}
\begin{align}
\gamma_{2}^{\rm E}(k, a) &= \frac{m^2(a)\frac{dm^2(a)}{da}\beta^2(a) \Omega_m}{2H_0^4\Pi(k,a)\Pi(k_1,a)\Pi(k_2,a)}\frac{k^2}{a^4H^2}\nonumber\\
&= -\frac{R\beta_0^2\Omega_m}{\xi_0^4}\frac{k^2}{a^{5+4R}H^2\Pi(k,a)\Pi(k_1,a)\Pi(k_2,a)}\exp^{\frac{2S}{2R-3}\left( a^{2R-3} - 1 \right)} \,\comma
\end{align}
where
\begin{align}
\beta(a) = \beta_0 \exp^{\frac{S}{2R-3}\left( a^{2R-3} - 1 \right)}\,\comma
\end{align}
\begin{align}
m(a) = \frac{H_0}{\xi_0} a^{-R}\,\period
\end{align}

\section{Comparison to SPT and alternative schemes for modeling the non-linear neutrino density}\label{sect:compnu}

In this appendix we show a comparison of our code with linear theory and standard perturbation theory (SPT). The SPT results were obtained by following the method of \cite{Saito:2009ah} using the Einstein-de Sitter approximation. We have also carried out a test using an alternative scheme to include the neutrino density in the Poisson equation Eq.~(\ref{eq:nlpoissoneq}). In this scheme we use
\begin{align}
\delta_\nu = \delta_{cb}\frac{\delta_\nu^{\rm lin}}{\delta_{cb}^{\rm lin}}\, \comma
\end{align}
where $\delta_{cb}$ is the non-linear CDM+baryon density contrast instead of $\delta_\nu = \delta_\nu^{\rm lin}$ in Eq.~(\ref{eq:nlpoissoneq}). This is what \cite{Blas} calls the improved external source scheme.

\begin{figure*}
\includegraphics[width=0.9\columnwidth]{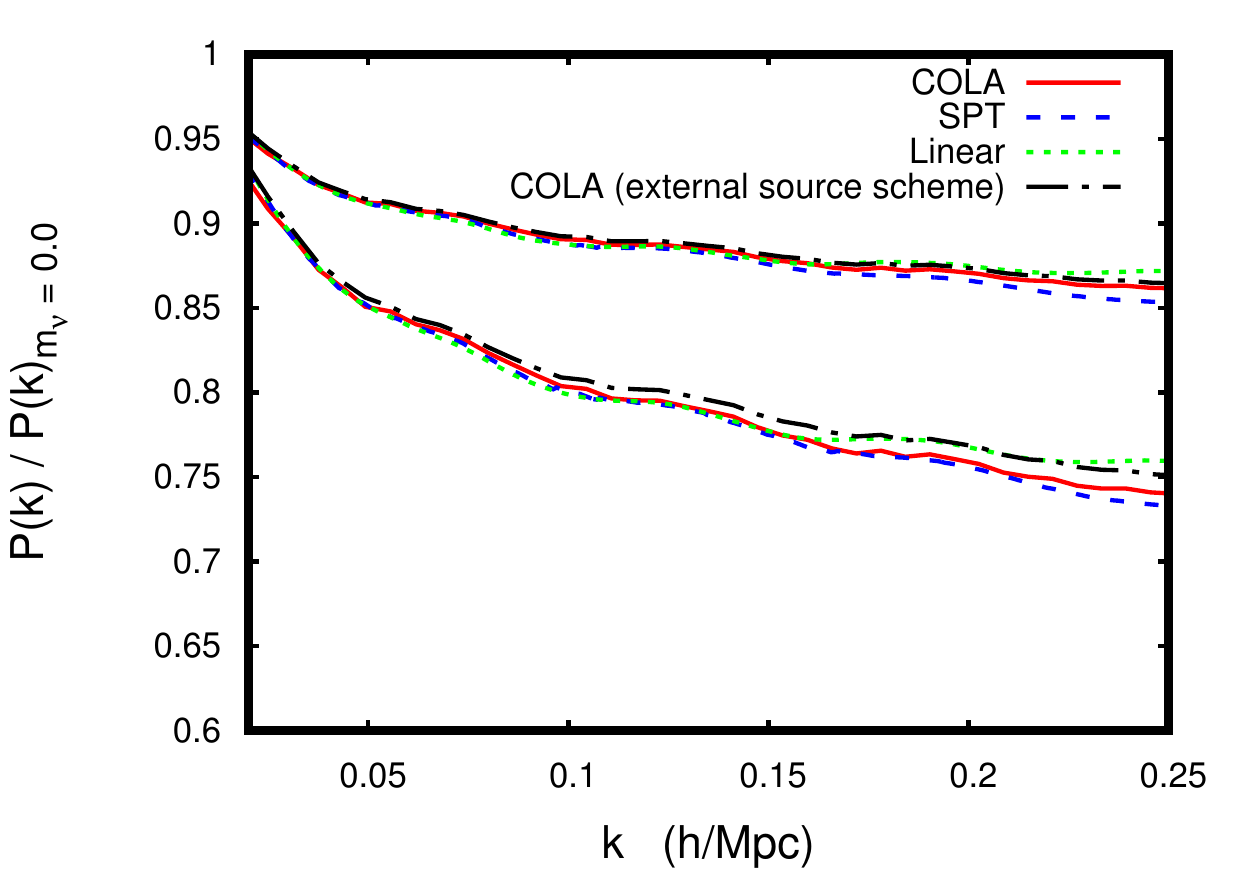}
\includegraphics[width=0.9\columnwidth]{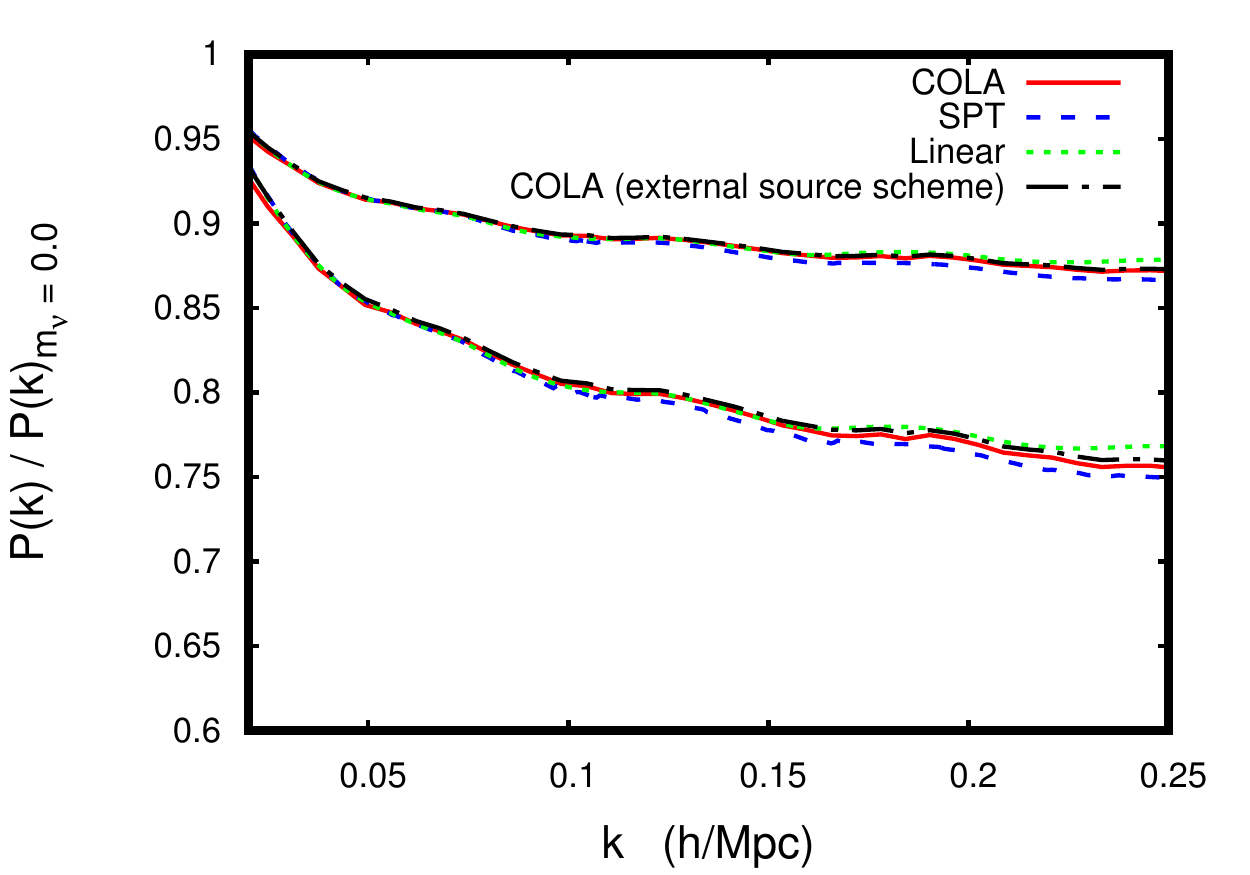}
\caption{The total matter power-spectrum for a GR+$m_{\nu}$ cosmology relative to the GR case where $m_\nu = 0.0$ at $z=0.0$ (above) and $z=1.0$ (below). We show the results of a COLA run compared to linear theory, SPT and what we get when we use the external source scheme in COLA. The upper lines in each figure shows the results for $m_\nu = 0.2$ eV and the lower lines shows the results for $m_\nu = 0.4$ eV.}
\label{fig:comparison}
\end{figure*}

Fig.~(\ref{fig:comparison}) shows that our implementation gives a result for the total matter power-spectrum that lies between SPT and linear theory on quasi-linear scales. The differences we see with respect to SPT for the slightly larger wavenumbers is expected (see Fig.~(10) in \cite{Blas}) as SPT slightly underestimates the power on these scales. The alternative scheme we tested is seen to slightly overestimate the power on linear scales and generally performs a bit worse on linear scales and at low redshift than the scheme we are using in this paper, especially for larger values of the neutrino mass.

Additionally, \cite{Blas} warns of inaccuracies at large scales when treating neutrinos in a purely linear way with $\delta_\nu = \delta_\nu^{\rm lin}$, where there is a large deviation from the full non-linear scheme at large scales as a consequence of violation of momentum conservation. Indeed such inaccuracies would appear from Eqs.~(\ref{eq:2LPTLCDMmnufull}) and (\ref{eq:2LPTMGmnufull}), but our method does not suffer from them because the approximations we make to maintain the speed of our COLA approach in the final pair of equations in Sections \ref{ssec:2ndorderLCDMmnu} and \ref{ssec:2ndorderMGmnu} demand that $D_{2, \rm cb}(\vec{k}, \vec{k}_1, \vec{k}_2, \tau) = \left( 1 - \frac{(\vec{k}_1 \cdot \vec{k}_2)^2}{k_1^2 k_2^2} \right) \hat{D}_{2, \rm cb}(k, \tau) \rightarrow 0$ as $k \rightarrow 0$.

\end{appendix}

\end{document}